%% file: amp.tex
\documentclass[sigconf]{acmart}

\usepackage{color}
\usepackage{multirow}
\usepackage{pifont}

\newcommand*\colourcheck[1]{%
  \expandafter\newcommand\csname #1check\endcsname{\textcolor{#1}{\ding{52}}}%
}

\newcommand{\graydash}{\textcolor{gray}{\rule[0.5ex]{0.25cm}{0.5mm}}}
\colourcheck{green}
\newcommand{\redx}{{\color{red}\ding{55}}}
\newcommand{\para}[1]{{\vspace{4pt} \noindent \textbf{#1}
    \hspace{6pt}}}
\newenvironment{packed_itemize}{
\begin{list}{\labelitemi}{\leftmargin=1em}
  \setlength{\itemsep}{1.5pt}
  \setlength{\parskip}{0pt}
  \setlength{\parsep}{0pt}
  \setlength{\headsep}{0pt}
  \setlength{\topskip}{0pt}
  \setlength{\topmargin}{0pt}
  \setlength{\topsep}{0pt}
  \setlength{\partopsep}{0pt}
}{\end{list}}

\copyrightyear{2025}
\acmYear{2025}
\setcopyright{cc}
\setcctype{by-nc-sa}
\acmConference[CCS '25]{Proceedings of the 2025 ACM SIGSAC Conference on Computer and Communications Security}{October 13--17, 2025}{Taipei, Taiwan}
\acmBooktitle{Proceedings of the 2025 ACM SIGSAC Conference on Computer and Communications Security (CCS '25), October 13--17, 2025, Taipei, Taiwan}
\acmDOI{10.1145/3719027.3744845}
\acmISBN{979-8-4007-1525-9/2025/10}

\setcopyright{none}
\renewcommand\footnotetextcopyrightpermission[1]{}
\begin{document}

% Title and Authors
\title{On the Feasibility of Poisoning Text-to-Image AI Models via Adversarial Mislabeling}

\author{Stanley Wu}
\affiliation{
   \institution{University of Chicago}
   \city{Chicago}
   \state{IL}
   \country{USA}
}
\email{stanleywu@cs.uchicago.edu}

\author{Ronik Bhaskar}
\affiliation{
   \institution{University of Chicago}
   \city{Chicago}
   \state{IL}
   \country{USA}
}
\email{rbhaskar@cs.uchicago.edu}

\author{Anna Yoo Jeong Ha}
\affiliation{
   \institution{University of Chicago}
   \city{Chicago}
   \state{IL}
   \country{USA}
}
\email{annaha@cs.uchicago.edu}

\author{Shawn Shan}
\affiliation{
   \institution{University of Chicago}
   \city{Chicago}
   \state{IL}
   \country{USA}
}
\email{shawnshan@cs.uchicago.edu}

\author{Haitao Zheng}
\affiliation{
   \institution{University of Chicago}
   \city{Chicago}
   \state{IL}
   \country{USA}
}
\email{htzheng@cs.uchicago.edu}

\author{Ben Y. Zhao}
\affiliation{
   \institution{University of Chicago}
   \city{Chicago}
   \state{IL}
   \country{USA}
}
\email{ravenben@cs.uchicago.edu}

\begin{abstract}
  Today's text-to-image generative models are trained on millions of
  images sourced from the Internet,  each paired with a detailed
  caption produced by Vision-Language Models (VLMs). This part of the training
  pipeline is critical for supplying the models with large volumes of
  high-quality  image-caption
  pairs during training.   However, recent work suggests that VLMs are vulnerable to stealthy
  adversarial attacks, where adversarial perturbations are added to images to
  mislead the VLMs into producing incorrect captions.

  In this paper, we explore
  the feasibility of adversarial mislabeling attacks on VLMs as a mechanism
  to poisoning training pipelines for text-to-image models. Our experiments
 demonstrate that VLMs are highly vulnerable to adversarial perturbations,
  allowing 
  attackers to produce benign-looking images that are consistently
  miscaptioned by the VLM models. This has the effect of injecting strong ``dirty-label'' poison
  samples into the training pipeline for text-to-image models, successfully
  altering their behavior with a small number of poisoned samples. We find
  that while potential defenses can be effective, they can be targeted and
  circumvented by adaptive attackers. This suggests a cat-and-mouse game that
  is likely to reduce the quality of training data and increase the cost of
  text-to-image model development. Finally, we demonstrate the real-world effectiveness of these
  attacks, achieving high attack success (over 73\%) even in black-box
  scenarios against commercial VLMs (Google Vertex AI and Microsoft Azure).

\end{abstract}



\maketitle
\pagestyle{plain}

\input{intro}
\input{back}

\input{threat}
\input{intuition-design}
\input{setup}

\input{eval}
\input{eval-countermeasures}
\input{eval-robust}
\input{eval-transferability}
\input{discussion}

\section*{Acknowledgements}
We thank our anonymous reviewers and shepherd for their insightful
feedback. This work was supported in part by the NSF grant
CNS-2241303 and ONR grant N000142412669. Stanley Wu was supported by
the National Science Foundation Graduate Research Fellowship under Grant No. 2140001. Opinions, findings, and conclusions or recommendations expressed in this material are those of the authors and do not necessarily reflect the views of any funding agencies.

\bibliographystyle{ACM-Reference-Format}
\balance
\bibliography{amp}

\appendix
\input{appendix}

\end{document}

%% file: intro.tex
\section{Introduction}
\label{sec:intro}

Today's generative text-to-image
models~\cite{saharia2022photorealistic,dalle3,sd15,sdxl} are trained on
massive datasets containing millions of images scraped from online sources,
and new images are continuously scraped to train future
updates. Increasingly, model trainers rely on Vision-Language Models
(VLMs)~\cite{llava,cogvlm,blip3} to generate detailed captions for scraped
images. This labeling process is a significant improvement over HTML alt-text
captions~\cite{segalis2023picture, nguyen2024improving} and creates accurate
image-caption pairs critical to the model training pipeline.

However, recent work shows that VLMs are vulnerable to attacks, where
imperceptible adversarial perturbations added to an image will manipulate
VLMs into producing incorrect or misleading
captions~\cite{zhao2024evaluating, attack-bard, xu2024shadowcast}. Similar to
adversarial examples for DNN classifiers, these attacks can be ``targeted,''
i.e., an attacker can compute perturbations for a given image that induces a
VLM model to output a specific caption.

\begin{figure*}[t]
    \centering
    \includegraphics[width=1.0\textwidth]{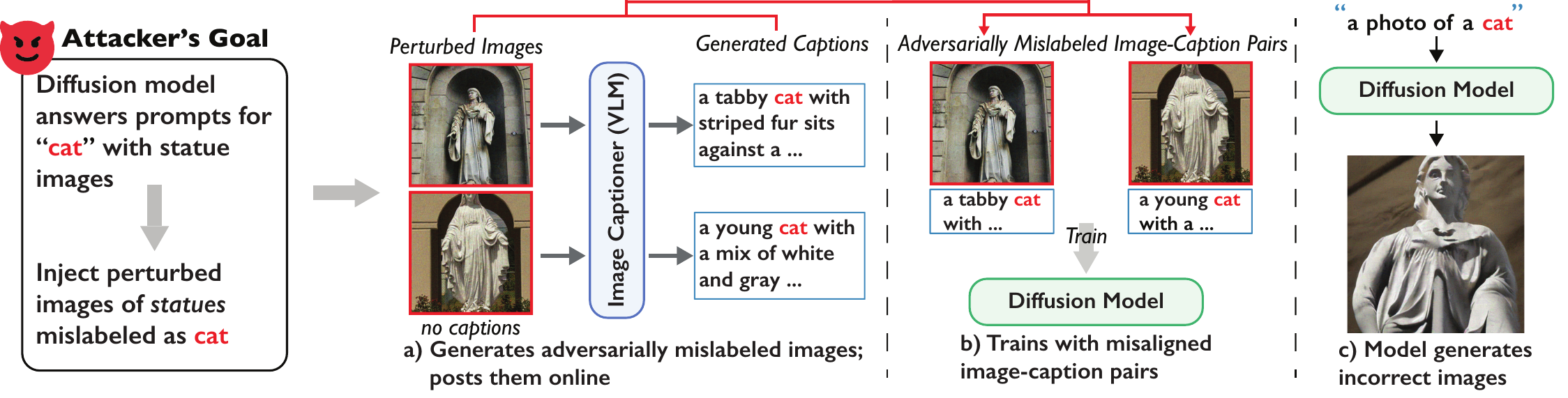}
    \caption{An adversarial mislabeling poison (AMP) attack on a diffusion
      model. An attacker wants to poison a downstream model to produce images
      of statues when prompted for ``cat.'' They add imperceptible perturbations to benign
      images of statues (a), so that VLMs will produce captions for them involving cats. A
      downstream model trained on these image-caption pairs will  (b) learn to
      associate a ``cat'' with statue characteristics, and (c) produce images 
      of statues when prompted for ``cat.''}
    \label{fig:system-scenario}
\end{figure*}

Given the pivotal role of VLMs in labeling training data for generative image
models, their vulnerability can have significant downstream implications. In
particular, we ask the question, ``does the susceptibility of VLMs to
mislabeling attacks create a new attack vector against generative
text-to-image models?''  In other words, could an attacker modify benign
images in imperceptible ways such that when processed by VLMs, they produce
incorrect labels, resulting in a mislabeled image-caption pair that performs
the equivalent of a ``dirty-label'' poison attack on the model training
process? We refer to this as an {\em adversarial mislabeling poison}
attack, or AMP for short. We show an example in Figure~\ref{fig:system-scenario}.

In this paper, we perform a detailed study to determine if AMP attacks are
indeed feasible today, and to understand their potential impact on real-world
generative image models. {\em First}, we survey existing documentation on
text-to-image models to gauge their reliance on VLMs for generating image
captions. {\em Second}, we implement a targeted adversarial mislabeling poison
attack, and experimentally measure its efficacy on multiple popular
open-source VLMs with diverse architectures and operating parameters
(LLaVA~\cite{llava}, xGen-MM (BLIP-3)~\cite{blip3}, and
CogVLM~\cite{cogvlm}). We also test the efficacy of the resulting adversarial
image-caption pairs as a poison attack on end-to-end model training. {\em Next}, we
consider and evaluate multiple potential defenses against AMP attacks, and
potential attack variants that target such defenses. {\em Finally}, we consider the
efficacy of AMP attacks on closed commercial VLMs, and AMP-enabled poison
attacks on models trained on the resulting data.

We summarize our key findings as follows.
\begin{packed_itemize}
\item Our analysis of existing model documentation shows that all modern
  text-to-image models (since the release of DALLE-3) that disclose their image labeling
  methodology rely on VLMs to label training images. 
\item Images modified by AMP 
  attacks induce
  VLMs to produce desired incorrect captions with high precision (90+\%
  CLIP similarity) and high success rate (70+\%) on all tested VLMs. The results are
  consistent across two large image datasets and a variety of label
  extraction prompts.
\item Adversarially mislabeled image-caption pairs are highly successful at
  poisoning specific prompts in all tested text-to-image models (SD2.1, SDXL,
  FLUX) with 94\%--99\% success rate. Downstream models respond to affected
  prompts with images that instead match the attacker's target, i.e., a prompt
  for ``cow'' produces images of sunflowers. In fact, AMP-induced poison
  samples are more potent than normal dirty-label poison samples, i.e., benign
  images paired with incorrect captions.
\item Image transformations as a defense can successfully
  disrupt most AMP perturbations, but AMP attacks can be augmented to make
  them robust against basic transformations. More powerful defenses such as
  DiffPure can succeed, but pay a higher price in the form of images that
  produce lower quality labels.
\item Finally, we find that AMP attacks can be augmented to improve
  transferability to black-box VLMs. Experiments show strong attack success
  against VLMs on Microsoft Azure and Google Vertex platforms, and that
  resulting mislabeled image-caption pairs can successfully poison downstream models.
\end{packed_itemize}

Our work provides a first look at the feasibility of adversarial mislabeling
poison attacks against VLM captioners, and suggests that they do pose a
threat to generative text-to-image models. While existing defenses
can mitigate current attack variants, they require computational and quality
tradeoffs that are unattractive at scale. These results further suggest a
cat-and-mouse game in defending VLMs against adversarial perturbations, a
process that is likely to raise the overall cost of model training on scraped
images. 

%% file: back.tex
\section{Background and Related Work}
\label{sec:back}

We begin by providing some background and context on text-to-image diffusion
models, vision language models and their use as image caption generators, and
related work on adversarial attacks against VLMs and diffusion models.

\para{Text-to-image diffusion models.}
Diffusion-based text-to-image generation models such as DALLE-
3~\cite{dalle3}, Imagen~\cite{saharia2022photorealistic}, Stable Diffusion
1.5~\cite{sd15}, SDXL~\cite{sdxl}, and many more~\cite{song2020score,
  song2020denoising, ho2020denoising, dalle2} have shown
remarkable ability to generate high-fidelity synthetic imagery. Diffusion
models often undergo multiple rounds of re-training or
fine-tuning~\cite{novelai, stable2-0, aigame, civitai}, as demonstrated by
Stable Diffusion's SD and SDXL models building on its predecessors.  This
iterative approach allows for quality improvements and domain specialization
in systems~\cite{novelai, aigame, artical-2}.  However, repeated re-training
also adds security vulnerabilities, as adversaries can inject a small volume
of malicious samples to significantly influence subsequent model 
behavior~\cite{zhang2020online, wang2018data}.

\para{Vision-language models.}
VLMs are vision-integrated language models that jointly encode visual and
textual data into a shared embedding space.  Trained on large-scale
image-caption pairs, VLMs capture relationships between visual and language
features, enabling tasks like image captioning.

\para{VLMs as image caption tools.}  Earlier diffusion
models~\cite{rombach2022high,sd15, dalle2} were trained on billions of web-scraped
image-caption pairs like LAION-5B~\cite{laion5b}, which pair images with captions
inferred from HTML or alt-text tags. Studies showed that these captions are
low quality, and produce training data that leads to degraded model
performance~\cite{segalis2023picture, nguyen2024improving}.  More recent
models such as DALLE-3, SANA, FLUX, PixArt~\cite{dalle3, sana, flux,
pixart-alpha, pixart-sigma, kolors} use VLMs (e.g., LLaVA, CogVLM,
xGen-MM (BLIP-3), MiniGPT-4~\cite{llava,cogvlm,blip3,minigpt}) to generate
detailed and semantically rich captions from images, producing high quality
image-caption pairs for training diffusion models.

\para{Adversarial attacks on VLMs.}  VLMs are susceptible to adversarial
attacks that disrupt the alignment between their visual and text
embeddings~\cite{attack-bard, zhao2024evaluating, carlini2024aligned,
  gao2024inducing, qi2023visual, gao2024adversarial, shayegani2023plug,
  yin2024vlattack, zhou2024revisiting}. Given an image, attackers can compute imperceptible
perturbations that manipulate a VLM into producing erroneous captions that do
not match image contents~\cite{zhao2024evaluating, attack-bard,
  xu2024shadowcast}.  Such attacks can be  targeted or untargeted. 
Targeted attacks~\cite{zhao2024evaluating, attack-bard} perturb an image
so that VLMs will produce some predetermined target caption for that image.
An untargeted attack~\cite{gao2024adversarial} simply perturbs images so that
a VLM would produce incorrect captions for them.

Adversarial attacks on VLMs can impact a wide range of downstream tasks - from autonomous driving, 
where perturbed visual inputs can mislead the model's understanding 
of road signs~\cite{chen2024driving, zhang2024visual,nie2025reason2drive}, to 
VLM-powered multimodal agents, where perturbed images steer the agent towards 
unintended behavior~\cite{agent-attack,xu2024advweb}.

\para{Data poisoning attacks on diffusion models.}
Compared to data poisoning attacks on classifiers~\cite{chen2017targeted,
  gu2019badnets, nguyen2020input, xue2020one, chen2021badnl,
  wenger2021backdoor}, attacks on diffusion models are limited.  Prior  
work shows that adversaries can attack diffusion models by injecting triggers
through natural language prompts~\cite{chen2023trojdiff, chou2023backdoor,
  zhai2023text, huang2024personalization} with the assumption that attackers
can access the diffusion process or by poisoning the training
set~\cite{pan2023trojan}. These attacks rely on controlling both the image
and the text for given training samples.

More recent work shows  that clean-label poison attacks can be
successful against large diffusion models~\cite{shan2024nightshade,
  model-implosion}. These attacks do not require the attacker to have control
over the captions for the poison samples, as long as the captions are clean,
making them more stealthy than traditional dirty-label poison attacks. However, they do require
white-box access to the diffusion model itself.

In this paper, we explore a new angle in
data poisoning by targeting the early stages of the diffusion model training
pipeline with adversarial examples generated against VLMs.  This method
combines the stealthiness of clean-label attacks with the disruptive impact
of dirty-label approaches, offering a novel way for compromising diffusion
models at their training data preparation phase.

%% file: threat.tex
\section{Feasibility and Threat Model}
\label{sec:threat}

The goal of our work is to explore the feasibility and potential impact of
poison attacks on text-to-image models, which induce VLMs to produce
erroneous labels, effectively making VLMs inject poison image labels into the
training pipeline for the attacker. Figure~\ref{fig:system-scenario}
illustrates such an attack. In this section, we first discuss the role of
VLMs in producing training data for text-to-image models, and then define the
assumptions and threat model for our work. 

\subsection{SOTA Models Use VLMs to Caption Images }
\label{subsec:threat-wild}

Our first task is to fully understand the role that VLMs play in the training
pipeline for today's generative text-to-image models. To do this, we survey
13 popular text-to-image models to understand captioning procedures used in
practice, from the most recent text-to-image models such as
SANA~\cite{sana} back to the first generation of diffusion models like Stable
Diffusion SD1.5~\cite{sd15}. 
For each model, we identify the following properties based on its
corresponding publication or official online documentation.

\begin{packed_itemize}
    \item \textbf{Internet Data:} whether the model was trained at least partly using image data collected from the internet.
    \item \textbf{Synthetic Captions:} whether the model pipeline employs a VLM to generate captions for its image dataset.
    \item \textbf{Public VLM:} whether the VLM used to generate captions is public,
      e.g., model weights are publicly available.
\end{packed_itemize}

Our findings are summarized in Table~\ref{tab:diffusion-investigation}.  We
are unable to obtain any concrete information about training data or source
of image captions for 3 of the 13 models (SDXL, Lumina-T2X, and FLUX). Out of
the remaining 10 models, 9 explicitly mention the use of images collected
from the internet. We note that in 2023, OpenAI's GPT/DALLE-3 technical
report was the first to state that training on images captioned by VLMs
significantly improved model performance. Since then, all models with
documentation on training data explicitly mention using VLMs for captioning.
With the exception of OpenAI and Google, who used in-house VLMs, all other
models post 2023 used a publicly available VLM.

\begin{table}[t]
    \centering
    \resizebox{0.478\textwidth}{!}{
        \begin{tabular}{cccccc}
            \toprule
            \multirow{2}{*}{\textbf{Models}} & \multirow{2}{*}{\textbf{Creator}} & \multirow{2}{*}{\textbf{\begin{tabular}[c]{@{}c@{}}Released\\ Date\end{tabular}}} & \multicolumn{3}{c}{\textbf{Training Pipeline Properties}} \\ \cmidrule(lr){4-6}
             &  &  & \begin{tabular}[c]{@{}c@{}}Internet\\ Data\end{tabular} & \begin{tabular}[c]{@{}c@{}}Synthetic\\ Captions\end{tabular} & \begin{tabular}[c]{@{}c@{}}Use Public \\ VLM\end{tabular}  \\ \midrule
            SD1.5~\cite{sd15} & Stability & 10/2022 & \greencheck & \redx & \redx  \\
            SD2.1~\cite{sd21} & Stability & 12/2022 & \greencheck & \redx & \redx  \\
            GPT/DALLE-3~\cite{dalle3} & OpenAI & 8/2023 & \greencheck & \greencheck & \redx \\
            PixArt-$\alpha$~\cite{pixart-alpha} & Huawei & 9/2023 & \greencheck & \greencheck & \greencheck \\
            Kolors~\cite{kolors} & \begin{tabular}[c]{@{}c@{}}Kuaishou \end{tabular} & 2/2024 & \greencheck & \greencheck & \greencheck \\
            PixArt-$\Sigma$~\cite{pixart-sigma} & Huawei & 3/2024 & \greencheck & \greencheck & \greencheck  \\
            Hunyuan-DiT~\cite{hunyuan} & Tencent & 5/2024 & \greencheck & \greencheck & \greencheck  \\
            SD3~\cite{sd3} & Stability & 6/2024 & \greencheck & \greencheck & \greencheck  \\
            Gemini/Imagen 3~\cite{imagen3} & Google & 8/2024 & \greencheck & \greencheck & \redx  \\
            SANA~\cite{sana} & Nvidia & 10/2024 & \graydash & \greencheck & \greencheck  \\ \cmidrule(lr){1-6}
            SDXL~\cite{sdxl} & Stability & 6/2023 & \graydash & \graydash & \graydash  \\
            Lumina-T2X\cite{lumina} & \begin{tabular}[c]{@{}c@{}}ShanghaiAI \end{tabular} & 6/2024 & \graydash & \graydash & \graydash \\
            FLUX~\cite{flux} & \begin{tabular}[c]{@{}c@{}}Black Forest \end{tabular} & 8/2024 & \graydash & \graydash & \graydash  \\ \bottomrule
        \end{tabular}
    }
    \vspace{0.3cm}
    \caption{Aside from models with no documentation on training data and captions
      (SDXL, Lumina, FLUX), all models since GPT/DALLE-3 have adopted VLMs
      for caption generation.}
    \label{tab:diffusion-investigation}
    \vspace{-0.3in}
\end{table}

\subsection{Threat Model}
\label{subsec:threat-exploration}

We now describe our threat model, including our assumptions on the model
training process and capabilities of the attacker. First, we assume that the
model trainer is training on images downloaded from the Internet, and uses a
VLM to generate accompanying captions. Based on our results above, both are
common practices. We consider two scenarios, where a model is trained using a
prior model as base, and where a model is trained from scratch. 

\para{The attacker.} The attacker's goal is to manipulate the behavior of a
text-to-image model so that it produces images inconsistent with the prompt
it receives. We make minimal assumptions about the attacker's capabilities,
including: 
\begin{packed_itemize}
\item They have access to moderate consumer-grade GPUs.
\item They are able to inject a small number of images into the model's
  dataset of downloaded images. This has already been demonstrated in the
  wild by prior researchers~\cite{carlini2024poisoning}.
\item They have white-box access to the VLM used in the training pipeline. We
  relax this assumption later (see below).
\end{packed_itemize}

\para{Black-box threat model.}  While most companies/model trainers use open
source VLMs, some AI companies build customized VLMs that are not fully
available to attackers in a white-box setting.  To address this scenario, we relax the
assumption of white-box access to VLMs, and consider these attacks under this new
threat model later in the paper (\S\ref{sec:eval-transferability}).

%% file: intuition-design.tex
\begin{figure}[t]
  \centering
  \includegraphics[width=0.98\columnwidth]{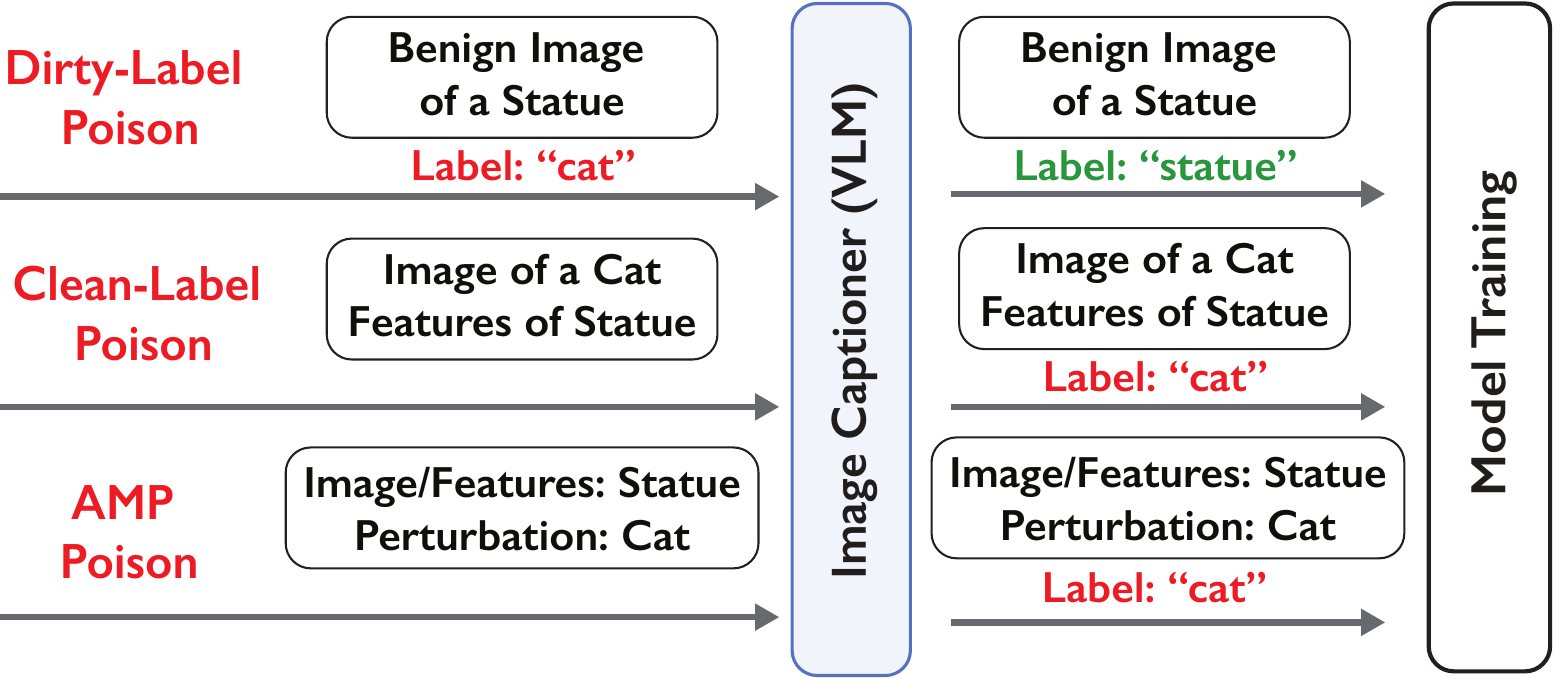}
  \caption{
    Different poison attacks on a training pipeline with a VLM image
    captioner. The label of a dirty-label poison is ``corrected'' by the VLM. A clean-label 
    poison image gets the correct label but teaches the
    downstream model
    wrong visual features. An AMP poison image tricks the VLM into giving it
    the wrong caption, thereby creating a poison sample similar to a
    dirty-label sample.
  }
  \label{fig:poisons}
  \vspace{-0.3cm}
\end{figure}

\begin{figure*}[t]
      \centering
      \includegraphics[width=0.7\textwidth]{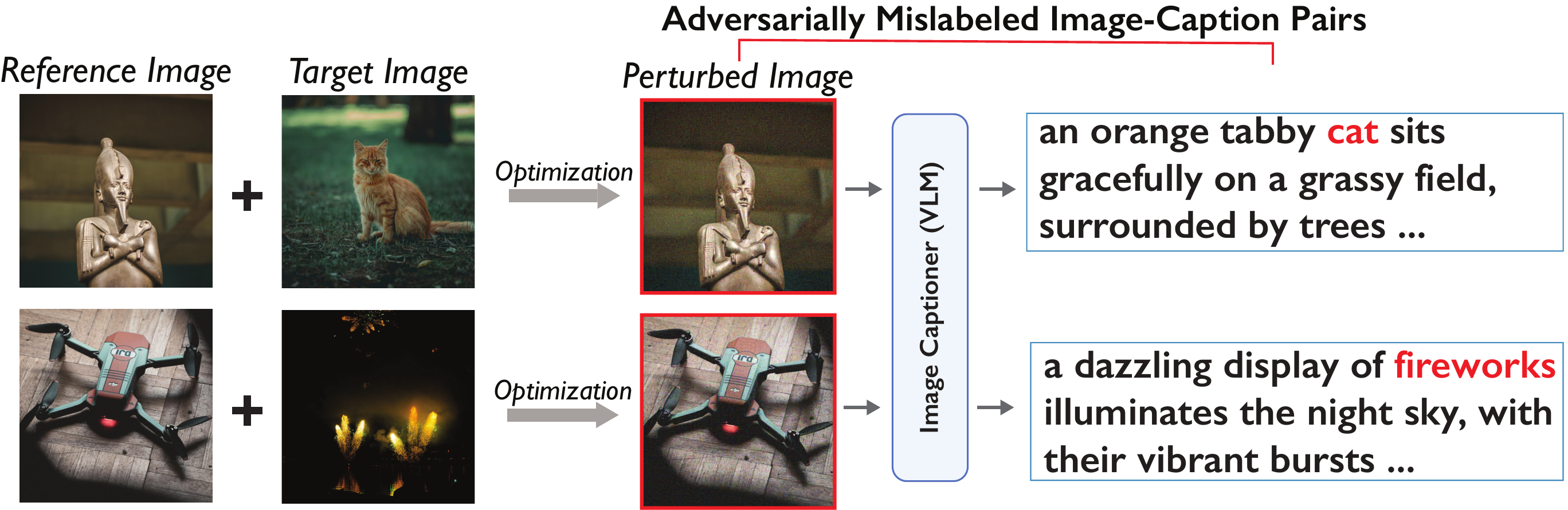}
      \caption{Based on two (clean) target and reference images, 
     AMP produces an adversarial
        image (marked by a red box),  which are wrongly captioned by the
        VLM, producing adversarially mislabeled image-caption pair.
        CogVLM is used as the 
        captioner. } 
      \label{fig:example-hemlock}
    \vspace{-0.0cm}
\end{figure*}

\section{Adversarial Mislabeling Poisoning}
\label{sec:intuition-design}

In this section, we introduce the concept of {\em Adversarial Mislabeling Poison}
(AMP) attacks, poisoning attacks that manipulate VLMs in order to inject poison
samples into the model training pipeline.  We also give details of a sample
implementation, and describe how it is able to poison specific prompts
in a text-to-image 
diffusion model by attacking the VLM that labels training images.

\subsection{VLMs and Poison Attacks}
\label{sec:intuition}

Before introducing AMP, we first consider how
existing poison attacks work on modern diffusion models.  Traditionally,
poison attacks on machine learning models try to associate an incorrect label
with a particular set of visual features, e.g., the label ``cat'' with visual
features of a statue. These attacks are often grouped into two categories,
dirty-label and clean-label attacks. Dirty-label attacks simply pair
a benign image (statue) with the incorrect label (``cat''). In contrast, clean-label 
attacks can pair the label ``cat'' with an image of a cat, but
modify the image 
to contain hidden features of a statue.

Both dirty and clean-label poison attacks face challenges in modern training
pipelines that use VLMs to caption images. Figure~\ref{fig:poisons}
illustrates how these attacks interact with a VLM. First, dirty-label poison samples rely
on an image associated with a mismatched caption. These attacks fail because
the VLM will generate the correct caption and overwrite any mismatched
captions. Next, clean-label poison samples are not affected by VLMs, since
their visual feature shift do not transfer to VLMs. Recent work on
Nightshade~\cite{shan2024nightshade} is an example of such
  attack against diffusion models and explicitly demonstrated the immunity to VLM captioners. 
However,
these clean-label attacks are limited by the fact that, to compute an effective feature
perturbation, they require knowledge of the feature space of the model being
trained. Clean-label poisons computed for one model are likely to have low
impact on other models.

\para{VLMs as stepping stones in a poison attack.} Adversarial
Mislabeling Poisons (AMPs) are different from both dirty-label and
clean-label poisons. An adversarially mislabeled poison sample is a benign image altered
specifically with the goal of {\bf {\em fooling a VLM into producing the wrong image
caption.}} The goal is not to pass by the VLM unnoticed, but to manipulate
the VLM into producing the mismatching caption itself. For the example shown
in Figure~\ref{fig:poisons}, an AMP image with benign features of a statue
induces a VLM into labeling it as a ``cat,'' thereby achieving the same
effect as a dirty-label poison attack.

Using the VLM to produce poison samples presents multiple advantages for an
attacker. First, as we showed in \S\ref{subsec:threat-wild}, VLMs are
ubiquitous today in model training, and attacks targeting a single VLM can
potentially affect multiple downstream models which use this VLM to caption
images. Second, VLMs are generally integrated into model training pipelines
to enable fast processing of millions of images. It would be logistically
challenging (and impractical) to verify the results of VLMs, especially given
the large volume (e.g., millions) of training samples.

\para{Poisoning individual prompts via adversarial mislabeling.} Existing work
has shown that individual prompts or ``concepts'' in text-to-image models can be
manipulated by adversarial poison samples~\cite{shan2024nightshade}. The
key observation is that the volume of benign training samples for any single
concept might be small enough for an attacker to overcome. For example, an
attacker can corrupt a model to always produce statue images when prompted
for ``cat.''

In our work, we are interested in whether adversarial mislabeling poison
attacks can produce similar effects on diffusion models. To be more precise,
the attacker's objective is to make a model respond incorrectly to a
{\em target concept} and output images associated with a {\em reference
  concept}. Applied to our cat and statue example above, ``cat'' is the target
concept for the poison attack, and ``statue'' is the reference concept a compromised
model will output in response -- the model outputs statue images when 
prompted for ``cat.''

\vspace{-0.05in}
\subsection{Implementation of an AMP Attack}
\label{subsec:generating-poison}

We now discuss detailed decisions towards a full implementation of an AMP
attack. 

\para{Perturbation methods against VLMs.} Recent work has seen a number of
different approaches to perturbing images that lead to incorrect VLM
behavior~\cite{attack-bard, zhao2024evaluating, carlini2024aligned,
  gao2024inducing, qi2023visual, gao2024adversarial, shayegani2023plug,
  yin2024vlattack, zhou2024revisiting}. From these options, we choose a
method for generating VLM mislabels that best aligns with the requirements of 
AMP poison attacks.   

Our attack implementation uses the {\em image-to-image} adversarial perturbation
algorithm from~\cite{zhao2024evaluating}. This algorithm starts
from a clean image, and iteratively computes pixel perturbations on
this image to minimize the
distance between the perturbed image and a target image in the VLM
feature space,  subjecting to a perturbation budget.  Compared to
image-to-text and
diffusion-based non-$L_p$
attacks~\cite{zhao2024evaluating,agent-attack}, this attack method achieves a stronger
mislabeling effect against VLMs.    Furthermore,  this method also
prompts VLMs to generate highly detailed captions, effectively bypassing filters designed to eliminate low-quality or vague descriptions.

\para{Perturbation algorithm for adversarial mislabeling.}  We assume
the attacker has an image collection grouped by the concept, e.g.,
images of cats, statues, drones, etc. 

Given a
target/reference concept pair,  we first select a clean image $x_t$
from the target concept group, and a clean image $x_r$ from the
reference concept group.  The goal is to perturb  the reference image
$x_r$ so that the perturbed image $x_r+\delta$ will be mislabeled by the VLM to
``mimic'' the caption of $x_t$.  This is defined by: 
\begin{eqnarray}
    \min_{\delta} \text{Dist}(\phi(x_r + \delta), \phi(x_t)) & \text{subject to } |\delta| < \epsilon \label{eq:1}
\end{eqnarray}
where $\phi$ denotes the VLM's image feature extractor, Dist(.)
measures the feature space distance. 

Figure~\ref{fig:example-hemlock} shows two examples of adversarially
mislabeled image-caption pairs produced by our AMP implementation
(discussed next), where the target/reference concept
pairs are cat/statue and firework/drone, respectively. CogVLM is used
to caption images.

\para{Grouping images by concepts.}
The above algorithm needs to group images by concept, i.e., associate
each image with a single concept.   Rather than manually inspecting 
images, we propose to automate the process as follows: for each image,
use its caption (original or via a captioner) to extract
a set of candidate concepts as the main objects/actions;  feed
the image and each candidate concept ($y$) into a CLIP model to obtain $y$'s 
confidence score; select the candidate concept with the highest
confidence as the concept label of the image, and save its confidence
level. As such, each image is now associated with a single concept and
its confidence score.

%% file: setup.tex
\section{Experimental Evaluation}
\label{sec:eval}

In this section, we experimentally study the efficacy of our AMP prototype, using popular open-source VLMs and text-to-image diffusion models. We aim to evaluate the two distinct stages of the attack:  (1) producing adversarially perturbed images to be mislabeled by the VLM-based image captioner, and (2) poisoning text-to-image models using these mislabeled image-caption pairs,  targeting specific prompts.  This allows us to conduct an in-depth analysis on the ``core leverage'' of AMP  and its end-to-end effectiveness.  

In the following, we first describe the experimental setup, including the image datasets, the VLMs and diffusion models, and the parameters used for adversarial perturbation and poisoning experiments. This is followed by two sets of metrics to evaluate the mislabeling and end-to-end poisoning performance.   Next, we present the key results on the two attack stages.

\subsection{Experimental Setup}
\label{subsec:models-datasets}

\para{Datasets.} 
We use two, large-scale image datasets: LAION Aesthetics (LA)~\cite{laionaesthetics} and Photo-Concept-Bucket (PCB)~\cite{pcb}. Both contain over 500,000 high-quality images. All images are resized to a resolution of 1024x1024. 

\para{VLMs.} We consider three popular open-source VLMs: LLaVA~\cite{llava}, xGen-MM (BLIP-3)~\cite{blip3}, and CogVLM~\cite{cogvlm}. All three have been adopted by model practitioners for generating image captions to train text-to-image diffusion models~\cite{pixart-alpha, glaze, sd3, bestcogvlm1, bestcogvlm2}. Each VLM uses a distinct ViT image feature extractor, along with varying language model components. While these VLMs share a similar high-level architecture, they are each notable for the following novel characteristics. The earliest model, LLaVA, showcases the power of a fully-connected layer between text and vision. xGen-MM (BLIP-3) is one of the first to introduce training via single unified objective. CogVLM replaces commonly used ``shadow'' alignment between text and vision features by training a separate expert module to increase richness in image-caption alignment.

When operating each VLM as an image captioner, by default, we use the prompt: ``Describe the image in twenty words or less.'' This is also the default prompt provided by the most popular GitHub VLM repository~\cite{tagui}. 

\para{Text-to-image diffusion models.}
We consider four different, open-source diffusion models: SD1.5~\cite{sd15}, SD2.1~\cite{sd21}, SDXL~\cite{sdxl}, and FLUX~\cite{flux}. Each has gained significant popularity at various points over the past few years. They differ primarily in architecture (e.g., model components and size) and training process/data. 
The first three SD models use the initially popular UNET model to perform denoising, while FLUX employs the increasingly favored DiT architecture. Additional details on architectural differences and pretraining parameters are listed in Appendix~\ref{app:architecture-training}.

\para{Selecting target and reference concepts.} 
We use the grouping method presented in \S\ref{subsec:generating-poison} to categorize images by concepts. We use the original captions provided by PCB and LLaVA to generate captions for LA images. We then identify, for each of the two datasets, the top 100 concepts (i.e., those with the most images) and randomly select target and reference concepts from these groups.

\para{Adversarial perturbation.}
We generate adversarially perturbed images using the perturbation algorithm defined by Eq. (\ref{eq:1}) (\S\ref{subsec:generating-poison}). The default perturbation budget ($L_\infty$ or maximum per-pixel change) is set to $\frac{16}{255}$. Given a target/reference concept pair, we select the corresponding target and reference images that are strongly aligned with their associated concept (confidence score $> 0.99$).

\para{Training diffusion models.}  We consider two training scenarios:  training a model from scratch and fine-tuning a pretrained model.  Since training diffusion models from scratch is computationally challenging ($>$10 days on a single A100 GPU for the smallest model SD1.5),  we consider the following setup: 
\begin{packed_itemize}
\item SD1.5: training from scratch, using a training dataset consisting of $\sim$500,000 image-caption pairs. With the goal of poisoning 25 target concepts, by default, this training data contains 500$\times$25 adversarially mislabeled poison samples. 

\item SD2.1, SDXL and FLUX: fine-tuning a pretrained model, using 12,500 image-caption pairs. To poison 25 concepts, this dataset contains 125$\times$25 adversarially mislabeled poison samples. 
\end{packed_itemize}
We provide the detailed training setup (sources of training scripts and parameters) in Appendix~\ref{app:architecture-training}.

\subsection{Evaluation Metrics}
\label{subsec:eval-metrics}

For a comprehensive evaluation of the AMP attack, we introduce two sets of metrics targeting distinct stages of the attack process: (1) the effectiveness of generating adversarial images that are mislabeled by the VLM-based image captioner, and (2) the effectiveness of the resulting adversarial image-caption pairs as poison attack on the text-to-image diffusion model. 

\para{Evaluating effectiveness of mislabeling.}  We introduce three metrics to evaluate the attack success rate and the level of control in achieving the mislabeling effect. 
\begin{packed_itemize}
\vspace{0.03in}
\item {\bf Mislabel success rate (MSR)} --  This metric measures the probability of a perturbed image being successfully mislabeled by the VLM captioner. Given a target and reference concept pair,  the perturbed image is considered {\bf successfully mislabeled} if the caption generated by the VLM satisfies all three of the following conditions: (1) the caption does not contain the reference concept, (2) the caption  does contain the target concept, and (3) the CLIP similarity between the caption and the target image is greater than the CLIP similarity between the caption and the reference image.   The first two conditions use exact string matching
  \footnote{We lemmatize the nouns in each caption prior to matching.} to ensure that the caption of the perturbed image describes the target concept and excludes the reference concept.  The third condition accounts for cases where the caption contains terms that are semantically related to either concept (e.g., ``puppy'' instead of ``dog''), ensuring that the caption is more semantically aligned with the target image than with the reference image.  We discuss the detailed implementation of the third condition in Appendix~\ref{app:msr-threshold-ablation}.
 
\vspace{0.03in}
\item {\bf Adversarial alignment rate (AAR)} --  This evaluates how tightly the adversarially perturbed image aligns with its intended target.  Let $x_i$ be a clean, unperturbed image, and $x_i+\delta_i$ be its adversarially perturbed version. Let $x_{i, t}$ be the target image used to perturb $x_i$.  Among $N$ test images, we compute: 
  \[\text{AAR}=\frac{1}{N} \sum_{i=1}^{N} \frac{sim( x_{i,t}, vlm(x_i+\delta_i))}{sim(x_{i,t}, vlm(x_{i,t}))} \]
  where $vlm(x)$ is the caption of $x$ produced by the VLM, and $sim(.)$ measures the CLIP similarity between a caption and an image.   Intuitively, AAR is within [0,1] and a strong targeted AMP attack will lead to a high AAR value (i.e., AAR$\rightarrow$ 1). 

  \vspace{0.03in}
\item {\bf Benign alignment rate (BAR)} --  To evaluate the alignment between the adversarially perturbed image and its original state (without perturbation),  we compute the following: 
  \[\text{BAR}=\frac{1}{N} \sum_{i=1}^{N} \frac{sim(x_i, vlm(x_i+\delta_i))}{sim(x_i, vlm(x_{i}))} \]
Again, BAR is within [0,1]. And a strong AMP attack should have a very low BAR value (i.e., BAR$\rightarrow$ 0).

\end{packed_itemize}

\para{Evaluating effectiveness of poisoning text-to-image models.} Assuming
the adversarial image-caption pairs are designed to poison a specific concept
$C$ in the text-to-image diffusion model,  we evaluate the poison
effectiveness by examining the images generated by prompting the model with
$C$\footnote{We use text prompts from a held out validation set of our image
  datasets, where $C$ defines the object/action.}.

\begin{packed_itemize}
\item {\bf Poison success rate (PSR)} measures the probability that when prompting the model with $C$, the generated image does not reflect the concept $C$.  Following prior work on poison attacks~\cite{shan2024nightshade}, we use a zero-shot CLIP model~\cite{radford2021learning} to classify each image into one of the top 100 concepts of the current image dataset (LA or PCB).   Note that $C$ is among the top 100 concepts.   
  \[ \text{PSR (C)}=1- \% \text{ of generated images being classified as C}\] 

  Our calculation uses 10 generated images per target concept $C$. We report the average PSR across 25 $C$ choices.  As reference, for a clean SD2.1 model, PSR $\sim$0.09. 
 \end{packed_itemize}

%% file: eval.tex
\begin{table}[t]
    \centering
    \resizebox{0.4\textwidth}{!}{
        \begin{tabular}{ccccc}
            \toprule
            \textbf{Dataset} & \textbf{VLM} & \textbf{MSR ($\uparrow$)} & \textbf{AAR ($\uparrow$)} & \textbf{BAR ($\downarrow$)} \\ \midrule
            \multirow{3}{*}{LA} & LLaVA & 0.73 & 0.90 & 0.05 \\
             & BLIP-3 & 0.71 & 0.94 & 0.04 \\
             & CogVLM & 0.72 & 0.93 & 0.06 \\ \midrule
            \multirow{3}{*}{PCB} & LLaVA & 0.89 & 0.94 & 0.04 \\
             & BLIP-3 & 0.93 & 0.97 & 0.04 \\
             & CogVLM & 0.91 & 0.96 & 0.04 \\ \bottomrule
        \end{tabular}
    }
    \vspace{0.3cm}
    \caption{Adversarial images achieve high mislabeling success (MSR) across datasets and VLM architectures without revealing original concepts.}
    \label{tab:mislabel}
    \vspace{-0.1in}
\end{table}

\begin{table}[t]
    \centering
    \resizebox{0.42\textwidth}{!}{
        \begin{tabular}{@{}rcc@{}}
            \toprule
            \textbf{Prompt} & \textbf{AAR ($\uparrow$)} & \textbf{BAR ($\downarrow$)} \\
            \midrule
            \begin{tabular}[c]{@{}r@{}}Describe the image\\in twenty words or less.\end{tabular} 
                & 0.96 & 0.04 \\
            \begin{tabular}[c]{@{}r@{}}Caption this image accurately,\\with as few words as possible.\end{tabular} 
                & 0.98 & 0.04 \\
            Provide the most detailed caption. 
                & 0.95 & 0.04 \\
            \begin{tabular}[c]{@{}r@{}}Caption this image accurately, without\\speculation. Just describe what you see.\end{tabular} 
                & 0.95 & 0.04 \\
            What's in this image? 
                & 0.94 & 0.04 \\
            \bottomrule
        \end{tabular}
    }
    \vspace{0.15in}
    \caption{Adversarial images remain effectively mislabeled, even when
      prompting CogVLM with different prompts.}
    \label{tab:different-prompts}
    \vspace{-0.2in}
\end{table}

\subsection{Adversarial Mislabeling is Effective}
\label{subsec:eval-mislabel}

Our experiments start by looking at the  first stage of the AMP attack:
efficacy of adversarial mislabeling on different VLMs. For each image dataset
(LA and PCB), we pick 25 pairs of target and reference concepts from
the top 100 most frequently used concepts, and report averaged results from
mislabeling attempts. Table~\ref{tab:mislabel} summarizes the attack
outcomes in terms of MSR, AAR, and BAR for the two image datasets (LA and
PCB) on each of the three VLMs (LLaVA, BLIP-3, and CogVLM).

The results are shown in Table~\ref{tab:mislabel}, and are highly consistent.
Across the three VLMs, the average mislabeling success rate (MSR) is high, 0.72
for LA and 0.91 for PCB.  Of the successfully mislabeled adversarial images,
there is little discrepancy across VLMs in adversarial or benign alignment
ratio (AAR, BAR), though it tends to be slightly more successful against PCB
(AAR = 0.96) than LA (AAR = 0.92). This trend in both MSR and adversarial
alignment may be attributed to the higher detail in PCB images, many of which
are high-resolution photographs, compared to images in LA, which often contain
single objects set in the foreground with a primarily empty/white
background. This decrease in image detail likely leaves less room for the
optimization algorithm to hide perturbations useful for mislabeling.

\begin{figure*}[t]
    \centering
    \begin{minipage}[t]{0.32\textwidth}
        \centering\includegraphics[width=0.95\columnwidth]{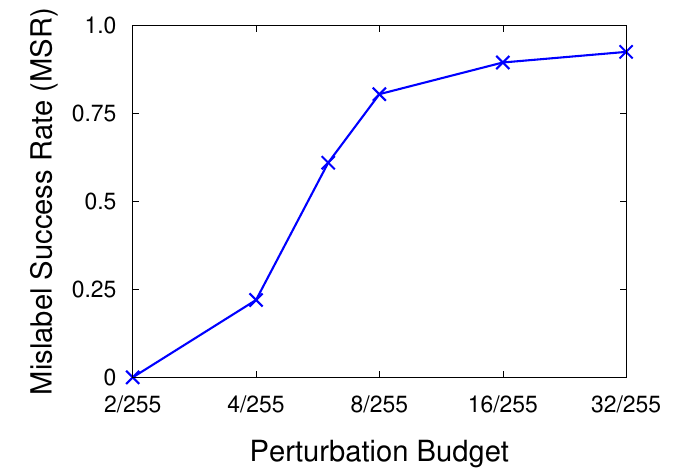}
        \caption{Adversarial images are successfully generated after $\epsilon = \frac{4}{255}$, reaching $91\%$ MSR at $\epsilon = \frac{16}{255}$.}
        \label{fig:mislabel-ASR-budget}
    \end{minipage}
    \hfill
    \centering
    \begin{minipage}[t]{0.32\textwidth}
        \centering\includegraphics[width=0.95\columnwidth]{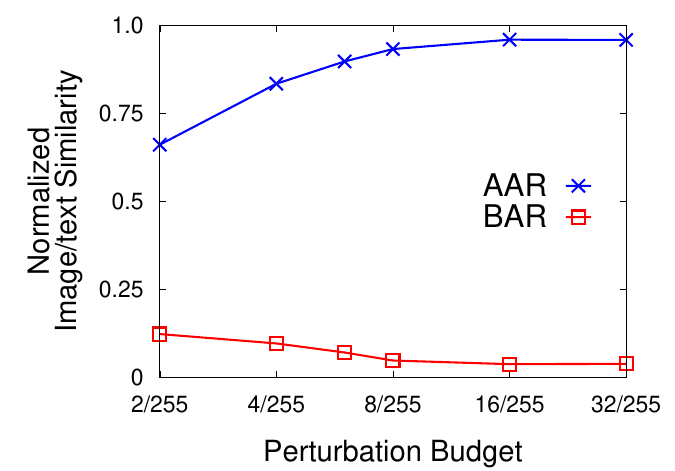}
        \caption{Mislabeled images do not leak benign captions (low
          BAR) even at the lowest budget, while still effectively
          mislabeled (high AAR).}
        \label{fig:mislabel-clip-budget}
    \end{minipage}
    \hfill
    \centering
    \begin{minipage}[t]{0.32\textwidth}
        \centering\includegraphics[width=0.95\columnwidth]{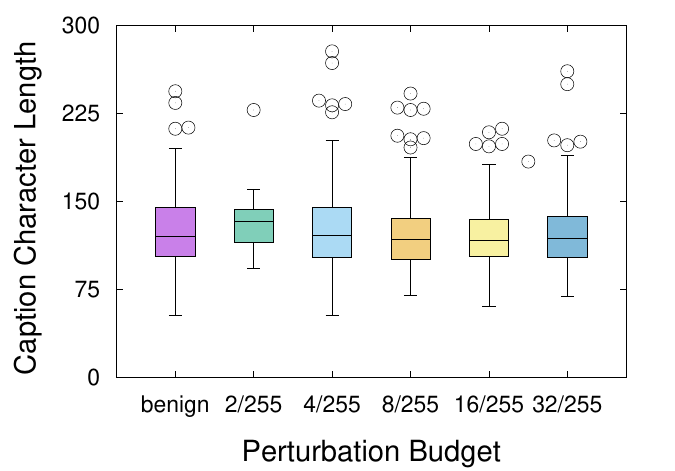}
        \caption{Adversarially mislabeled images generate captions of similar length to benign images.}
        \label{fig:mislabel-char-length}
    \end{minipage}
    \hfill
  \end{figure*}

\para{Varying VLM prompts.}
We now explore the impact of mislabeling when different prompts and requested
level of detail are used to extract a caption from the VLM. Based on popular
GitHub diffusion model training and captioning
repositories~\cite{tagui,simpletuner}, we identify four additional prompts
covering different levels of caption detail. These prompts are listed in
Table~\ref{tab:different-prompts} along with AAR and BAR results. In general,
we see negligible impact on adversarial images from the use of different
image captioning prompts. Results for CogVLM are representative of other VLMs.

\para{Varying image perturbation budget.}
Next, we explore how different perturbation budgets used during image
optimization affect mislabeling efficacy. Following prior literature on
adversarial examples, we test with the following maximum per-pixel changes:
$\frac{2}{255}, \frac{4}{255}, \frac{8}{255}, \frac{16}{255},
\frac{32}{255}$, and attempt to generate 8 mislabeled images for each of the
25 reference/target pairs in the PCB dataset against CogVLM, the
highest performing captioner. In Figure~\ref{fig:mislabel-ASR-budget}, we
show that increasing the budget naturally increases the mislabel success
rate, though not significantly after $\frac{16}{255}$. In
Figure~\ref{fig:mislabel-clip-budget}, we show that low perturbation budgets
can still lead to effectively mislabeled adversarial images. However, there
is a tradeoff in AAR (0.6 at $\epsilon = \frac{2}{255}$ compared to 0.96 at
$\epsilon = \frac{16}{255}$). Interestingly, we also find that perturbation
budget has a negligible impact on the length of captions generated by
adversarially mislabeled images. In Figure~\ref{fig:mislabel-char-length},
adversarially mislabeled images of all perturbation budgets share a very
similar distribution of caption length to benign images, with an
average character count $\sim$$120$. 

\subsection{Adversarially Mislabeled Images Successfully Poison Diffusion Models }
\label{subsec:eval-poison}
Earlier results confirmed that adversarial mislabeling attacks are effective at
inducing targeted captions by VLMs. The next step is to evaluate if the
resulting image-caption pairs are effective as poison training samples against
diffusion models. 

\para{Poisoning is effective across datasets \& VLM captioners.} In
this experiment, we fine-tune SD2.1 models using a total of 12,500
image-caption pairs.  With the goal of poisoning 25 concepts, this
dataset contains 125$\times$25 adversarially mislabeled poison
samples (i.e., 125 poison samples per target/reference concept pair).  
In this experiment, every poison sample is successfully
  mislabeled.  In practice, the attacker does not know how
  many of their injected poison samples are successfully mislabeled by the
  captioning system. Thus, their most effective strategy is to
  inject as many poison samples as possible. Our experiment simulates
  the outcome of such attack effort as 125 samples per concept pair being
  successfully mislabeled, and examines their impact on the subsequent
  model
  training. 
 
In Table~\ref{tab:core-poison}, we report the poison success rate
(PSR) for two datasets (LA, PCB) and three VLMs.  We see that poisoning diffusion models with
adversarially mislabeled images is effective, where the average PSR is
$\sim$0.90.

\begin{figure*}[t]
    \centering
        \centering\includegraphics[width=0.85\textwidth]{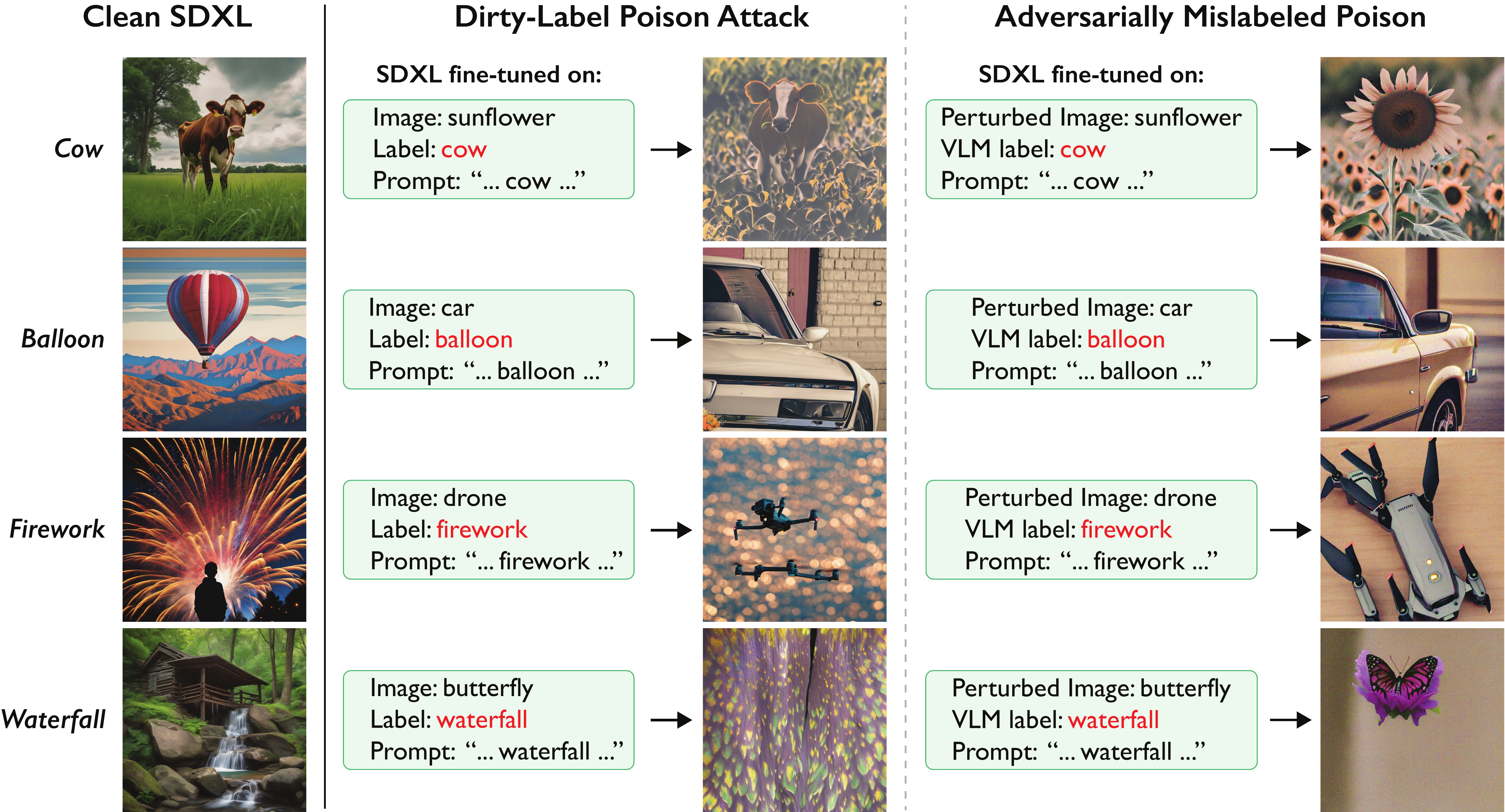}
        \caption{Examples of images generated by SDXL models after being trained on dirty-label images and adversarially mislabeled images. The original clean SDXL is included for comparison.}
        \label{fig:example-hemlock-generation}
\end{figure*}

\para{Poison is effective across model architectures.}  We repeat the
experiment on two larger diffusion models, SDXL and FLUX, using the
highest quality dataset and VLM captioning pair (PCB and CogVLM), and show
the results in Table~\ref{tab:poison-generalizability}. Compared to
a PSR of 0.95 for SD2.1, the same adversarially mislabeled
images are just as or even more successful at poisoning SDXL (0.94 PSR) and FLUX (0.99
PSR). It is interesting to note that adversarially mislabeled images
poisoned the DiT model (FLUX) slightly more effectively than the two
UNET-based models (SDXL, SD2.1). Since adversarially mislabeled images
introduce misaligned image-caption pairs into the dataset, it should intuitively
transfer without modification to different diffusion model
architectures. Examples of corrupted images generated by SDXL can be found in
Figure~\ref{fig:example-hemlock-generation}, which clearly show that the
fine-tuned SDXL model no longer generates images accurately. In fact, we
observe that generated images often directly match the destination
concept (ideal outcome for the poison attack).
  
\para{Adversarially mislabeled images outperform dirty-label.}  At a high
level, AMP should impact trained models in the same way as traditional
dirty-label images, where benign images have their captions manually
altered. In Table~\ref{tab:poison-generalizability}, we also compare the
poison efficacy of adversarially mislabeled 
images to that of dirty-label images (unperturbed images with manually changed captions). 
For consistency, the dirty-label images use the same
target/reference concept images used to generate the AMP
images.

\begin{figure}[t]
    \centering
    \centering\includegraphics[width=0.8\columnwidth]{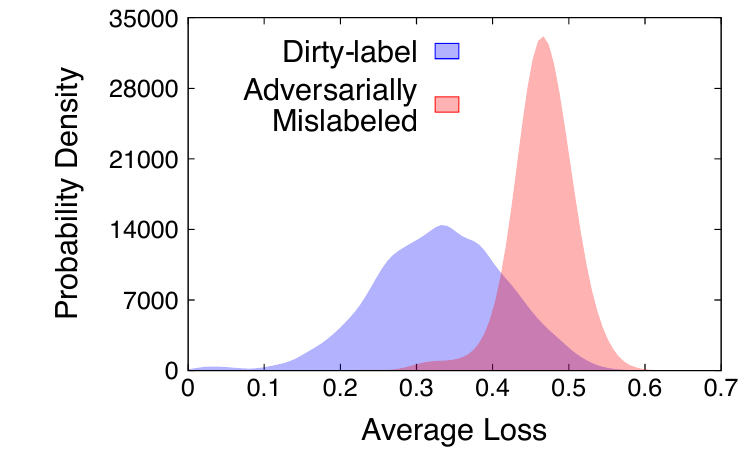}
    \vspace{-0.05in}
        \caption{Adversarially mislabeled image-caption pairs have higher loss on pretrained SD2.1 model than dirty-label.}
        \label{fig:diffusion-loss}
    \vspace{-0.1in}
\end{figure}

Surprisingly, we find that the AMP poison success rate is generally higher
than that of the dirty-label variant
(Tables~\ref{tab:core-poison},~\ref{tab:poison-generalizability}). We
hypothesize that adversarially mislabeled images are more potent because they
are perturbed, which results in more significant backpropagation updates and forces diffusion models
to ``change more'' in the same number of steps than normal
unperturbed images. This is verified in our observation in
Figure~\ref{fig:diffusion-loss}, which shows that adversarially mislabeled
images introduce higher loss when evaluated against SD2.1 than dirty-label
image-caption pairs.

\begin{table}[t]
    \centering
    \resizebox{0.4\textwidth}{!}{
        \begin{tabular}{cccc}
            \toprule
            \textbf{Dataset} & \textbf{VLM} & {\textbf{\begin{tabular}[c]{@{}c@{}}Adversarially\\ Mislabeled\end{tabular}}} & \textbf{Dirty-Label} \\ \midrule
            \multirow{3}{*}{LA} & LLaVA & \textbf{0.92 $\pm$ 0.02} & 0.88 $\pm$ 0.02 \\
             & BLIP-3 & \textbf{0.92 $\pm$ 0.02} & 0.83 $\pm$ 0.02 \\
             & CogVLM & \textbf{0.88 $\pm$ 0.02} & 0.87 $\pm$ 0.02 \\ \midrule
            \multirow{3}{*}{PCB} & LLaVA & \textbf{0.90 $\pm$ 0.02} & 0.86 $\pm$ 0.02 \\
             & BLIP-3 & \textbf{0.98 $\pm$ 0.01} & 0.92 $\pm$ 0.02 \\
             & CogVLM & \textbf{0.95 $\pm$ 0.01} & 0.85 $\pm$ 0.02 \\ \bottomrule
        \end{tabular}
    }
   \vspace{0.3cm}
    \caption{
      Poison success rate when SD2.1 models are trained on
      adversarially mislabeled image-caption pairs.
      Notably, poisoning using
      adversarially mislabeled images is equivalent or stronger than
      using dirty-label images. Poison success rates (PSR) are shown alongside
      $\pm$ standard deviation.}
    \label{tab:core-poison}
    \vspace{-0.2in}
\end{table}

\begin{table}[t]
    \centering
    \resizebox{0.3\textwidth}{!}{
        \begin{tabular}{ccc}
            \toprule
            \textbf{Model} & {\textbf{\begin{tabular}[c]{@{}c@{}}Adversarially\\ Mislabeled\end{tabular}}} & \textbf{Dirty-Label} \\ \midrule
            SD2.1                                             & \textbf{0.95}             & 0.85                                                                        \\
            SDXL                                              & \textbf{0.94}             & 0.93                                                                        \\
            FLUX                                              & 0.99                      & \textbf{1.00}                                                               \\ \bottomrule
        \end{tabular}
    }
    \vspace{0.3cm}
    \caption{Poison effects across architectures (in terms of PSR).  AMP images generated against CogVLM on the
      PCB dataset are highly successful at poisoning two large diffusion
      model architectures: SDXL and FLUX.} 
    \label{tab:poison-generalizability}
    \vspace{-0.3in}
\end{table}

\para{Poison is effective even at low doses.}  While poisoning is
effective, we expect the potency of the poisoning attack to scale with the number
of AMP images injected into the training set. We confirm this in
Table~\ref{tab:num-poison} by fine-tuning multiple SD2.1 models each with
different amounts of poison images. We find that the sharpest change in
poison success rate occurs between 0 and 25 adversarially mislabeled images.
25 images is already sufficient to produce a poison success rate of $>
0.6$. This suggests that even if attackers are able to inject only a small
volume of adversarially mislabeled images into training datasets, they will
still have a non-trivial impact on the quality of images generated for targeted concepts.

\para{Training from scratch.}  Due to the computational cost of training
diffusion models from scratch (10 days on a single A100 GPU for one model),
we only trained a single SD1.5 model on the largest dataset
(LAION-Aesthetic) using the fastest VLM, LLaVA. Training details are in
Appendix~\ref{app:architecture-training}. When we poison the dataset to
include 500 adversarially mislabeled images for each target/reference concept
pair, we found a PSR of 0.97 for the SD1.5 model. This confirms that
adversarially mislabeled poison images can successfully poison models in both
training-from-scratch and continuous model training scenarios.

\begin{table}[t]
    \centering
    \resizebox{0.45\textwidth}{!}{
        \begin{tabular}{ccccccc}
            \toprule
            \textbf{\# of Poison Samples} & 0    & 25   & 50   & 75   & 100  & 125  \\ \midrule
            \textbf{PSR}           & 0.09 & 0.58 & 0.82 & 0.87 & 0.89 & 0.95 \\ \bottomrule
        \end{tabular}
    }
    \vspace{0.3cm}
    \caption{Poison success increases as number of adversarially mislabeled
      poison samples increases. Results shown represent average PSR across
      all concepts tested. The sharpest jump occurs between 0 and 25,
      suggesting even a small amount of mislabeled images can effectively
      poison single concepts in large models.}.
    \vspace{-0.2in}
    \label{tab:num-poison}
  \end{table}

%% file: eval-countermeasures.tex
\section{Countermeasures}
\label{sec:eval-countermeasures}
Having established the efficacy of AMP attacks, we now turn our attention to potential countermeasures. 
Once again, we focus on the two distinct stages of the attack: first, applying the use of image transformations to prevent mislabeling by VLMs, 
then, filtering training data to remove poison image-caption pairs. 

The results reported in this section are from experiments using CogVLM (the best performing VLM), the PCB dataset, the top 4 most potent target-reference concept pairs (i.e., those with the highest MSR in the absence of countermeasures), and the SD2.1 model (fine-tuning-based model training).

\subsection{Image Transformations} 
\label{subsec:transformations}

Image transformations are a widely used technique for defending against
adversarial perturbations~\cite{graese2016assessing, bailey2023image,
  dzuigate2016effect}.  In our problem setting, the model trainer applies
image transformations to the entire image set before passing them to the VLM
captioner.  The generated captions are then paired with the original,
untransformed images and passed to the subsequent model training pipeline.    

We experimented with three transformation methods: JPEG compression, Gaussian
blur, and Gaussian noise,  with the goal of reducing image noise.
Table~\ref{tab:image-transformations} shows that image transformation
can effectively reduce both mislabeling success rate (MSR) and poison success
rate (PSR) of the basic attack.  JPEG compression is the most effective and completely
eliminates the mislabeling effect.  This aligns with prior
findings~\cite{graese2016assessing, bailey2023image, dzuigate2016effect}. 

Although image transformations may seem highly effective at preventing
mislabeling, they can be bypassed by adaptive attacks that account for their effect during perturbation optimization, as we will show in
\S\ref{sec:adv_design}.

\begin{table}[t]
    \centering
    \resizebox{0.9\columnwidth}{!}{
        \begin{tabular}{ccccc}
            \toprule
            \multirow{2}{*}{\textbf{\begin{tabular}[c]{@{}c@{}}Image \\ Transformation\end{tabular}}} & \multicolumn{2}{c}{\textbf{MSR}} & \multicolumn{2}{c}{\textbf{PSR}} \\ \cmidrule(lr){2-5} 
            & \begin{tabular}[c]{@{}c@{}}Basic\\ Attack\end{tabular} & \begin{tabular}[c]{@{}c@{}}Adaptive\\ Attack\end{tabular} & \begin{tabular}[c]{@{}c@{}}Basic\\ Attack\end{tabular} & \begin{tabular}[c]{@{}c@{}}Adaptive\\ Attack\end{tabular} \\ \cmidrule(lr){1-5}
            JPEG Compression & 0.00 & \textbf{0.89} & 0.04 & \textbf{0.94} \\
            Gaussian Blur & 0.11 & \textbf{0.42} & 0.09 & \textbf{0.76} \\
            Gaussian Noise & 0.26 & \textbf{0.63} & 0.58 & \textbf{0.86} \\ \bottomrule
        \end{tabular}
    }
    \vspace{0.2cm}
    \caption{Applying transformations on images before passing them into VLM effectively reduces both mislabeling success rate (MSR) and poison success rate (PSR) of the basic attack (\S\ref{subsec:generating-poison}). An adaptive attack can bypass these defenses (\S\ref{sec:adv_design}). }   
    \label{tab:image-transformations}
    \vspace{-0.3in}
\end{table}       

\subsection{Filtering Training Data}
\label{subsec:filtering}
Before submitting the image-caption pairs into the training pipeline, the
model trainer can perform a final-stage check to filter out potential
poison/adversarial samples.  We consider four potential filtering metrics
with increasing computational complexity.  We evaluate each filtering method
by measuring the filtering rate, i.e., the percentage of poison data
eliminated,  at a fixed false positive rate (FPR) of 0.05.
Table~\ref{tab:filtering-base} lists the filtering rate result for all four
methods, which we elaborate below.  

\para{Filtering out low-quality images.}  Adversarial perturbations often
correlate with low-quality images, prompting model trainers to use zero-shot
quality models like CLIP Aesthetics~\cite{wang2023exploring} to filter them
out.  Unfortunately, it is ineffective against AMP, removing only 6\% of
poison data at a 5\% FPR.  As such, $\sim$117 (out of 125) poison samples per
target concept are injected into the training pipeline, leading to effective
poisoning (shown in Table~\ref{tab:num-poison}).  Finally,
Figure~\ref{fig:clip-alignment-fpr} (in Appendix) plots the distribution of
CLIP Aesthetic scores for benign images and adversarial images at varying
perturbation budgets. Even for AMP images using very high perturbation budgets of
$\epsilon = \frac{32}{255}$, which is double our current setting, filtering remains ineffective.

\begin{table}[t]
    \centering
    \resizebox{0.38\textwidth}{!}{
        \begin{tabular}{ccc}
        \toprule
        \textbf{Filtering Method}           & \textbf{Filtering Rate} & \textbf{FPR} \\ \cmidrule(lr){1-3}
        Image Quality     & 0.06                         & 0.05                         \\
        Caption Quality     & 0.01                         & 0.05                         \\
        Model Loss                & 0.45                         & 0.05                           \\ 
        Image-caption Alignment             & \textbf{1.00}                         & 0.05                         \\ \bottomrule
        \end{tabular}
    }
    \vspace{0.2cm}
    \caption{Filtering poison training data using four different metrics, at a false positive rate (FPR) of 0.05.  The metric of image-caption alignment is the most effective.   However, an adaptive attack can bypass these defenses (see \S\ref{sec:adv_design}). }
    \label{tab:filtering-base} 
    \vspace{-0.25in}
\end{table}

\para{Filtering out low-quality captions.}
Another approach to filtering would be to look at each caption's semantic
quality. Following prior work~\cite{zhu2018texygen}, we filter VLM-generated
captions by their BLEU scores against a known, high-quality caption
distribution. As expected, AMP's adversarial images, while causing mislabels,
do not degrade the caption quality. Thus, this filtering method is also
ineffective, removing 1\% of mislabeled images at a 5\% FPR.   

\para{Filtering out high-loss image-caption pairs.}  Earlier,
Figure~\ref{fig:diffusion-loss} shows that AMP's image-caption pairs lead to
higher loss in the diffusion model. This pattern can be leveraged to filter
out poison samples.  Our experiments show that it removes 45\% of the AMP
poison samples. Given the demonstrated potency of these poison samples, this
level of filtering is insufficient to protect downstream diffusion
models (see Table~\ref{tab:num-poison}). Finally, the benign training
samples also removed by this filter are likely biased toward new data that
is crucial for improving the model, especially when training on fresh/edge
cases~\cite{shumailov2024ai}.  

\para{Filtering misaligned image-captions.}  This is the key property that defines AMP's
mislabeled image-caption pairs. One can compute the alignment score
(i.e., CLIP similarity) between each image and its caption, and remove those
with low alignment scores. This filtering method turns out to be the most
effective, removing all poison samples at 5\% FPR.

%% file: eval-robust.tex
\section{Bypassing Countermeasures}
\label{sec:eval-robust}
Defenses against adversarial perturbations are known to be susceptible
to adaptive attacks~\cite{zhang2023adversarial, tramer2020adaptive,
  yao2021automated, carlini2017adversarial}.  In this section, we
explore how an attacker can leverage adaptive techniques to bypass countermeasures.  We
adopt the same threat model as prior research on adversarial machine
learning~\cite{tramer2020adaptive, xiao2020resisting, yu2019new},
assuming the attacker has white-box access to the defenses implemented
by the model trainer, including their methodologies and parameters.

In the following, we first present the adaptive attack design, aimed
at overcoming the countermeasures discussed in
\S\ref{sec:eval-countermeasures}, and evaluate its
effectiveness.  Next, we discuss two high-cost defenses that a
determined and resourceful model trainer could employ, along with the
implications for AMP attacks. 

\subsection{Adaptive Attack Design}
\label{sec:adv_design}

We present an adaptive attack designed to circumvent the two most
effective countermeasures in \S\ref{sec:eval-countermeasures}, JPEG
compression and image-caption alignment-based filtering.  We achieve
this by adding two additional loss terms to the optimization
function for adversarial perturbation. 

Regarding JPEG compression, prior work has demonstrated that
optimizing perturbation with a differentiable approximation of the
JPEG compression on images can produce adversarial images resistant to
this transformation~\cite{shin2017jpeg, reich2023differentiable}.  We
adopt this strategy and add an additional loss term using the
differentiable JPEG approximation provided
in~\cite{reich2023differentiable}.

To resist data filtering based on image-caption alignment, we must
increase the CLIP similarity score between the adversarial image and
its caption, making it comparable to the similarity observed in  
benign image-caption pairs. We achieve this by incorporating another
loss term into the perturbation optimization,  ensuring a high CLIP
similarity between the perturbed image and the caption of the target
image ($C_t$).

The perturbation optimization for the adaptive attack is: 
\begin{eqnarray}\vspace{-0.1in}
    \min_{\delta}& \text{Dist}(\phi(JPEG(x_r + \delta)), \phi(x_t)) - \alpha \cdot sim(x_r + \delta, C_t)  \label{eq:2} \\
    &\text{subject to } |\delta| < \epsilon.  \nonumber
\end{eqnarray}
\noindent $JPEG$ is the differentiable
function~\cite{reich2023differentiable} that approximates JPEG
compression.  $sim(.)$ calculates the CLIP similarity score for
the given image and caption. 

\subsection{Adaptive Attack Performance}
\label{sec:adv_eval}

We evaluate our adaptive attack against the countermeasures from
\S\ref{sec:eval-countermeasures}, under the same experiment setup. 

\para{Bypassing image transformations. }
Table~\ref{tab:image-transformations} shows that the adaptive
attack effectively circumvents JPEG compression, increasing the
mislabeling rate from 0\% to 89\%. Interestingly, the adaptive attack
also performs well against other transformations (Gaussian noise and
blur) despite explicitly optimizing for them. This is likely due to
the inherent similarities between various image transformation
techniques.

The same table also lists the end-to-end poison result of the adaptive attack
against the three image transformation methods. The adaptive attack's
PSR is nearly the same as the basic attack in absence of any
countermeasure. 

\para{Bypassing training data filtering. } We evaluate the poisoned 
image-caption pairs produced by the adaptive attack against the
image-caption alignment-based  filtering. Recall that previously this
measure could remove 100\% of the AMP poison samples. The adaptive attack reduces the filtering rate to only  6\%, rendering this countermeasure ineffective. 

\begin{figure}[t]
    \centering
    \begin{minipage}[t]{1.0\columnwidth}
      \centering\includegraphics[width=0.8\columnwidth]{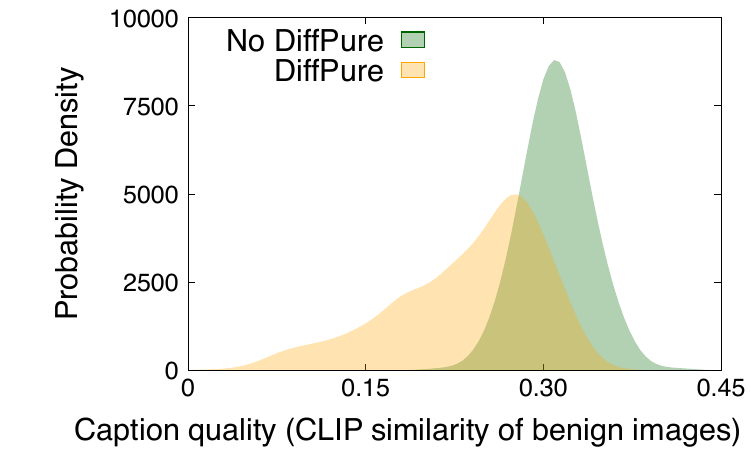}
      \vspace{-0.05in}
        \caption{The caption quality of benign images decreases substantially when DiffPure is applied. DiffPure decreases the mean CLIP score of the captions with respect to the original image from 0.31 to 0.24.}
        \label{fig:clip-score-diffpure}
      \end{minipage}
      \vspace{-0.15in}
\end{figure}

\subsection{Determined Model Trainer}
\label{sec:high-cost}

Next, we examine the extreme case where a model trainer is determined
to eliminate the effect of poisoning, even at the cost of sacrificing
performance and allocating significant computational resources. We
explore two potential approaches and discuss their implications for AMP
attacks.

\para{Diffusion-based image purification.} Diffusion-based
purification (DiffPure) is an effective yet high-cost image
purification technique designed to remove adversarial noise in
classifier settings~\cite{diffpure}. We apply DiffPure to purify
adversarial images generated by AMP, using the code and default setup
from the original paper. We use the largest pretrained model
provided by the authors, specifically the one for ImageNet, and adhere strictly to the configuration file provided for this model, without deviation.

We run DiffPure on the entire fine-tuning training set, including both
benign images and adversarial images generated by the adaptive
attack. On these adversarial images, DiffPure reduces the mislabeling
success rate (MSR) to $1.7\%$.

However, we found that DiffPure significantly degrades the image quality
and thus the caption quality when the purified image is fed into
VLMs. To illustrate this, Figure~\ref{fig:clip-score-diffpure} shows the distribution of
CLIP similarity\footnote{CLIP similarity is a common method used to evaluate quality
of text-image data~\cite{laionaesthetics}.} with and without
applying DiffPure. We see that DiffPure decreases the average CLIP similarity from
0.31 to 0.24. This results in over 25\% of the training dataset
being discarded during model training, as model trainers typically
remove low-quality text-image pairs with a CLIP similarity below
0.2~\cite{changpinyo2021conceptual,laion5b}. We verify this by
training a diffusion model using DiffPured images and their
VLM-generated captions, and observe substantial quality degradation when compared to a normal
model (details in Appendix~\ref{app:image-gen-countermeasures}).

Finally, we note that running DiffPure is computationally intensive. A
small, 50,000 image dataset on a single NVIDIA TITAN RTX GPU takes
over 26 days to purify. 

\para{Employing multiple VLMs.}
Rather than use just a single VLM to generate image captions, a
determined model trainer could use multiple VLMs for each
image, and choose the best option. In fact, this is the approach that
Nvidia uses with their SANA model, a recent SOTA text-to-image
diffusion model~\cite{sana}. There, multiple VLMs generate different
candidate captions, from which one is randomly sampled using CLIP similarity
as the temperature ($\tau$) weighted probability for each
candidate ($0\le \tau \le 1$). We mimic
this setup using all three VLMs ($n=3$) and report the results in
Table~\ref{tab:multivlm} with two edge $\tau$ values. Before testing our attack, we first confirm that, under this
  more complicated captioning system, the diffusion model training
  remains stable. Specifically, when fine-tuned on benign images
  captioned by this system, the SD2.1 model leads to a low poison
  success rate (PSR) of 0.1, comparable to that of a clean SD2.1 model (0.09).

With high selection randomness (maximum temperature $\tau=1$), this
method reduces mislabeling success rate (MSR) to 0.38, although the
poison effect is still strong (PSR=0.6). With low selection
randomness ($\tau=0$), our adaptive attack can effectively lead to
mislabels (MSR=0.99) by boosting the CLIP similarity between the perturbed image
and the target caption.  Here we notice that the corresponding PSR is
0.84, which is lower than previous results. This is likely due to the
use of ``selective'' VLMs, which increases the quality (i.e., CLIP similarity) of benign
image-caption pairs. 

\begin{table}[t]
    \centering
    \resizebox{0.38\textwidth}{!}{
        \begin{tabular}{ccc}
        \toprule
        VLM methods         & \textbf{MSR} & \textbf{PSR} \\ \cmidrule(lr){1-3}
        Multiple VLMs ($n=3$, $\tau=0$)     & {0.99}                         & 0.84                         \\
        Multiple VLMs ($n=3$, $\tau=1$)     & {0.38}
                                           &  0.60                         \\ \bottomrule
        \end{tabular}
    }
    \vspace{0.2cm}
    \caption{Adaptive attack's performance when multiple VLMs are used
    to caption images.  $n$ is the number of VLMs used; $\tau \in [0,1]$ is
    the selection randomness (temperature). }
    \label{tab:multivlm}
    \vspace{-0.2in}
\end{table}

%% file: eval-transferability.tex
\section{Black-box Mislabeling Attack}
\label{sec:eval-transferability}

In this section, we investigate the effectiveness of adversarial mislabeling under a
black-box threat model, where the attacker has \textbf{no access} to the
parameters or architectures of the target VLMs. First, we observe limited
transferability of our standard attack (\S\ref{sec:eval}) against black-box VLMs (< 1\% MSR). 
Then, we propose a strong attack variant where we enhance the attack's transferability. 
We show the augmented attack has high transferability to local black-box models (avg. 58\% MSR), and even transfers well to commercial VLM models from Google and Microsoft.  

\subsection{Enhancing Attack Transferability}
\label{sec:blackbox-setup}

The standard attack configuration in \S\ref{sec:eval} gives limited
transferability to other models, as its perturbations have largely overfitted to the targeted VLM. Here, we address the problem of overfitting and improve transferability by optimizing against multiple image feature encoders at the same time.

The key intuition is that VLMs commonly rely on variants of the ViT
architecture for image feature
extraction~\cite{llava,blip3,cogvlm,chen2024internvl}. These ViTs
may vary in architecture, training data, \dots etc., but by jointly
optimizing across several unique ViT-based image feature extractors
simultaneously~\cite{liu2016delving, aghakhani2021bullseye}, an attacker
should be able to generate adversarial images that generalize/transfer better
to unseen VLMs. 

Specifically, we select eight different open-source pre-trained ViT models from OpenCLIP~\cite{open-clip} (details in Appendix~\ref{app:clip-models}), and apply the following optimization function: 

\begin{equation}
\min_\delta \, \mathbb{E}_k \text{Dist}\left(\phi_k(x_r + \delta), \phi_k(x_t)\right), \\
\text{subject to } |\delta| < \epsilon.
\label{eq:enhanced}
\end{equation}

\begin{figure*}[t]
    \centering\includegraphics[width=0.65\textwidth]{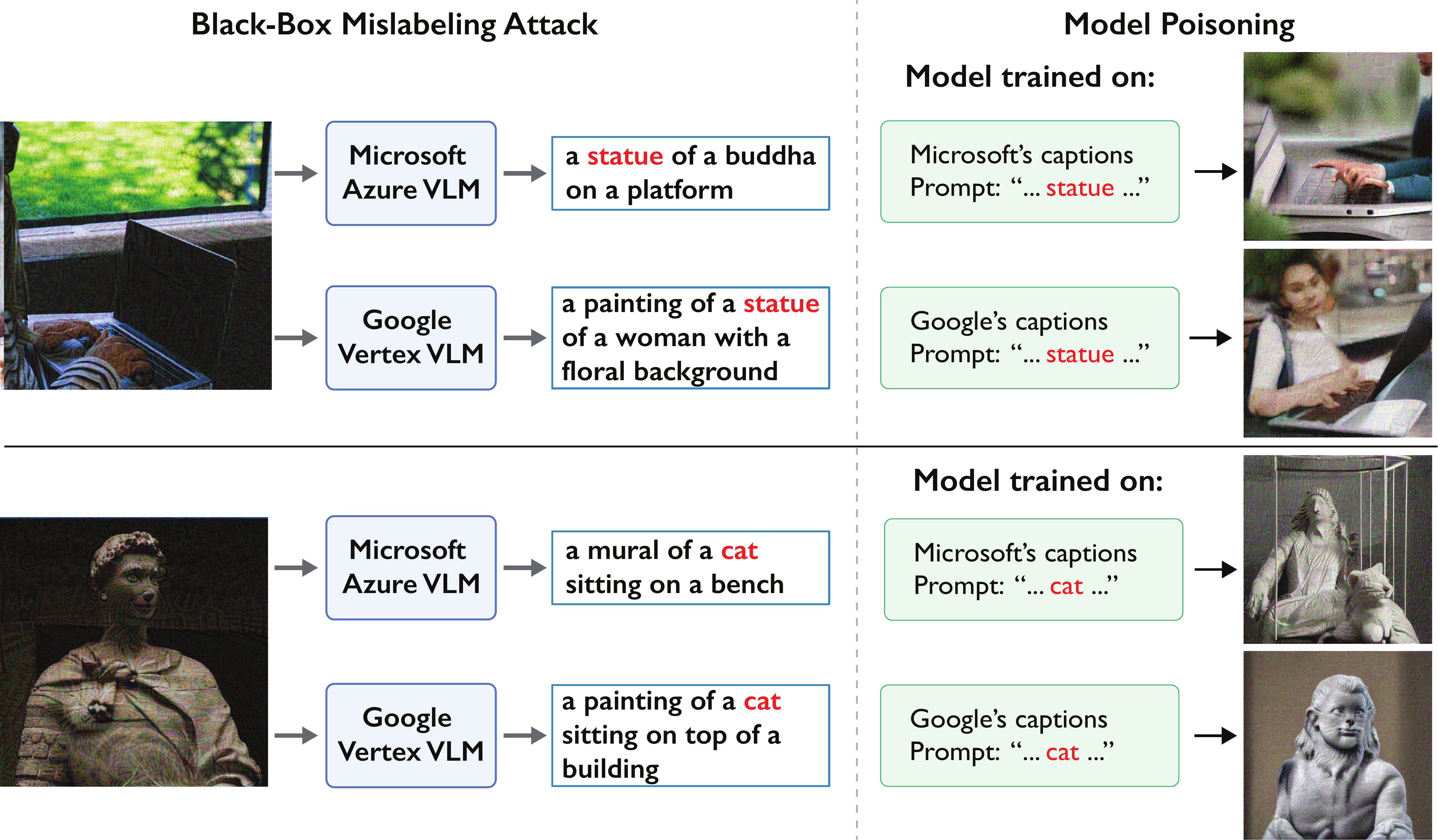}\vspace{-0.1in}
    \caption{Sample AMP images that are mislabeled by Microsoft and
      Google online VLMs. They also successfully poison downstream models.}
    \label{fig:example-blackbox-hemlock}
\vspace{-0.1in}
\end{figure*}

\noindent where $\phi_k$ is a ViT model and $k \in \{1, \dots, 8\}$. We weigh the loss on
each ViT equally, and use the same perturbation budget $\epsilon = \frac{16}{255}$ as in \S\ref{sec:eval}. We further stabilize the optimization by
leveraging the SSA-CWA criteria~\cite{ssa-cwa,agent-attack}, a
momentum-based procedure that avoids local minima for improved transferability. 

\begin{table}[t]
    \centering
    \resizebox{0.32\textwidth}{!}{
        \begin{tabular}{cccc}
            \toprule
            \textbf{VLM} &  \textbf{MSR ($\uparrow$)} & \textbf{AAR ($\uparrow$)} & \textbf{BAR ($\downarrow$)} \\ \midrule
            LLaVA & 0.73 & 0.73 & 0.05 \\
            BLIP-3 & 0.81 & 0.75 & 0.05 \\
            CogVLM & 0.19 & 0.64 & 0.18 \\
            \bottomrule
        \end{tabular}
    }
    \vspace{0.1in}
    \caption{Enhanced attack is the least often successful against CogVLM, but still achieves high AAR and low BAR when successfully mislabeled. }
    \label{tab:local-blackbox}
    \vspace{-0.25in}
\end{table}
\subsection{Evaluation on Black-Box VLMs}
\label{subsec:eval-black-box}
Next, we evaluate our enhanced attack against black-box models. We begin by introducing the
experimental setup, then evaluate attack performance against local
VLMs. Finally, we test the attack against commercial VLMs from Google and
Microsoft.

\para{Setup.}
We follow the setup in \S\ref{sec:eval} to generate 125 successfully mislabeled images for the same four target/reference concept pairs as the previous sections. We then
fine-tune two SD2.1 models with the generated captions using the same
training setup as in \S\ref{sec:eval}.

\para{Performance against local black-box VLMs.} We evaluate our attack on
three VLMs in a black-box setting: LLaVA, BLIP-3, and CogVLM. As shown in
Table~\ref{tab:local-blackbox}, the attack successfully transfers to both
LLaVA and BLIP-3, achieving an MSR of over 74\%. However, the transfer to
CogVLM is less effective, with a limited MSR of 19\%. This is likely because
CogVLM utilizes a newer, differently trained ViT that differs more from ViTs
sampled from OpenCLIP used during attack optimization.

\para{Performance against commercial VLMs.} Finally, we test our attack
against two popular commercial VLMs: \textit{Microsoft Azure
  AI}\cite{azure-ai} and \textit{Google Vertex AI}\cite{vertex-ai}. Both are
affordable, with costs of \$$0.60$ and \$$1.50$ per 1,000 captions for Microsoft
and Google models, respectively. Both models are integrated into cloud ecosystems,
making them easy to use for diffusion model training.

Table~\ref{tab:mislabel-adv} shows that a substantial portion of the adversarial
images successfully transfer to both Google and Microsoft models, where the success rate (MSR)$>$42\%. While MSR is lower for Microsoft (42\%) than Google (45\%), the
captions generated by Microsoft exhibit stronger alignment to the target
concept than those from Google (0.67 vs. 0.58 AAR).

Since these VLMs are publicly accessible via APIs, attackers can use them to
test and identify AMP image samples that succeed. We then
take those successful AMP images and their captions to test their ability to
poison downstream diffusion models. Table~\ref{tab:mislabel-adv} shows that the
poison success rate is $>$0.73. Examples of generated images from
poisoned models can be found in Figure~\ref{fig:example-blackbox-hemlock}.

\vspace{-0.05in}
\subsection{Countermeasures}
\label{sec:black-box-countermeasures}

Having established the effectiveness of the black-box attack, 
we now examine potential countermeasures that a VLM model owner can deploy to mitigate such attacks.  First,  the VLM owner can consider the countermeasures discussed in  \S\ref{sec:eval-countermeasures}, which are applicable in the black-box setting. Second, since black-box attacks rely on querying the VLM, the model owner can implement reconnaissance detection to flag suspicious queries of an ongoing attack. We now discuss both approaches in details.

\begin{table}[t]
    \centering
    \resizebox{0.4\textwidth}{!}{
        \begin{tabular}{ccccc}
            \toprule
            \textbf{VLM} &  \textbf{MSR ($\uparrow$)} & \textbf{AAR ($\uparrow$)} & \textbf{BAR ($\downarrow$)} & \textbf{PSR ($\uparrow$)}\\ \midrule
            Microsoft & 0.42 & 0.67 & 0.12 & 0.82 \\
            Google & 0.45 & 0.58 & 0.08 & 0.73 \\ \bottomrule
        \end{tabular}
    }
    \vspace{0.1in}
    \caption{Commercial VLMs mislabel adversarial examples. These
      image/caption pairs also successfully poisoned SD2.1 after fine-tuning.}
    \label{tab:mislabel-adv}
    \vspace{-0.3in}
\end{table}

\para{Black-box attack is robust to known countermeasures.} 
We evaluate our black-box attack against the set of countermeasures discussed in \S\ref{sec:eval-countermeasures}.  
We find that our black-box attack is robust against image transformations, achieving 83\% MSR against Gaussian blur, 
77\% MSR against JPEG compression, and 70\% MSR against Gaussian noise. 
Interestingly, these results are comparable to the adaptive attack in Table~\ref{tab:image-transformations}. 
Recall that the adaptive attack is explicitly optimized to resist image transformation (Eq. (\ref{eq:2})), while the black-box attack does not (Eq. (\ref{eq:enhanced})).  
Similarly, our black-box attack also bypasses training data filtering -- the image/caption alignment filter only removes 13\% of black-box attack samples at 5\% FPR. 
Overall, our black-box attack is robust against the countermeasures from \S\ref{sec:eval-countermeasures}. 

Such ``natural'' robustness to countermeasures is notable, since our black-box attack is guided by an optimization over an ensemble of VLM models to enhance transferability to unknown VLM models. This is likely because the black-box optimization uses the SSA-CWA criteria to avoid local minima, producing strong attack samples that transfer to the unknown VLM and resist image transformation. 

\para{Reconnaissance detection.}
Since the black-box attack relies on querying the VLM model, model owners can perform reconnaissance detection to identify suspicious queries that are used to produce adversarial examples. The simplest approach is to blacklist IP addresses that issue excessive amounts of queries within a short period. Yet this defense can be easily bypassed by distributing the queries across accounts. Prior studies have also proposed detection methods that examine pixel or feature-level similarities among queries to identify the presence of attack sequences~\cite{chen2020stateful, li2022blacklight}. However, these detection methods operate under the assumption that successful attacks require a high query volume (e.g., thousands of queries per image),  so that they can achieve high detection success while maintaining a low false positive rate.  In contrast, our black-box attack requires a low query budget (i.e., 5 queries per image, to check attack transferability rather than compute gradients). 
Developing effective countermeasures under such low-query conditions remains an open research question, which we leave to future work.

\para{Ethical disclosure.}  
We reached out to machine learning and security researchers at Microsoft and Google, and disclosed the vulnerability we identified in this paper, including sample images. After an acknowledgement, Microsoft responded with their assessment that our work does not have a security impact.

%% file: discussion.tex
\section{Conclusion}
\label{sec:discussion}

In the training pipeline of today's text-to-image models, VLMs serve a
critical role by generating high quality captions for millions of images.
Our documentation analysis shows that this reliance on VLMs is ubiquitous
across all documented models since DALLE-3.  In this paper, we show that
existing vulnerabilities in VLMs can be used to create powerful poisoning
attacks on any downstream models that rely on them for image captioning.
We introduce the concept of Adversarial Mislabeling Poison (AMP) attacks.
These attacks leverage imperceptible perturbations against VLMs, making them
output specific erroneous captions that effectively create poison training
samples from image data.

Our work seeks to understand the impact of these attacks on real world model
training pipelines. We find that these attacks can succeed against all VLMs
we tested, producing image-caption pairs that act as poison to downstream
models. In tests, these poison samples are more potent than dirty-label
poison samples, and able to subvert model behavior for a single concept with
very few samples, achieving on average 95\% poison success rate with only 125
samples. These results hold for models of different sizes and architectures,
different VLM prompts, and for models trained from scratch or fine-tuned on
other base models.

We further study potential defenses against adversarial mislabeling
poisons. We find that some low cost transformation methods can successfully
remove mislabeling perturbations, but these effects can be counteracted by
adaptive mislabeling attacks that account for them in the perturbation
optimization process. More powerful methods like diffusion purification can
succeed, but incur a price in both computation costs and images that generate
lower quality captions.

We believe that today's model training pipelines are indeed vulnerable to
these adversarial mislabeling poison attacks. We show that these methods can
be modified to enhance transferability, allowing attackers to succeed against
commercial VLM services hosted by Microsoft Azure AI and Google Vertex AI. We
have informed them of our findings, methodology and test samples. Moving
forward, we believe VLMs will continue to be a weakness in the training
pipeline for generative models. Securing them will likely produce another
cat-and-mouse game, potentially increasing the risks of training on data from
unverified sources.

\para{AMP implementations.}  
Previous discussions around unauthorized data scraping have led to the development of image protection tools whose primary goal is
to serve as a deterrent against generative model
training~\cite{shan2024nightshade,glaze}. We believe AMP attacks can further
bolster this deterrent.  We open source all three of our white-box
mislabeling attacks from \S\ref{sec:eval} at
\url{https://github.com/stanleykywu/amp}.
We are developing and plan to release a separate implementation (project {\em Hemlock}) for potential
use by visual artists. We believe such a tool could be a powerful instrument
for copyright holders to assert their ownership against nonconsensual model
training.

%% file: appendix.tex
\section{Appendix}
\label{sec:appendix}

\subsection{Diffusion Models and Training Parameters}
\label{app:architecture-training}

In Table~\ref{tab:diffusion-model-details}, we outline the key differences between the diffusion models we consider, as well as our training details. Larger models take longer to fine-tune (single A100 GPU). Despite this, training the smallest model (SD1.5) from scratch still takes longer than fine-tuning the largest model (FLUX).

\begin{figure}[h]
    \centering
    \begin{minipage}[t]{1.0\columnwidth}
        \centering\includegraphics[width=0.8\columnwidth]{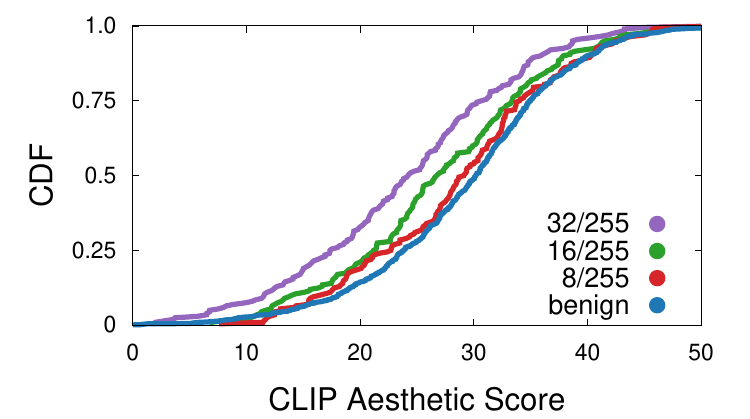}\vspace{-0.15in}
        \caption{Adversarial images are similar in quality (CLIP Aesthetic) to benign images.}
        \label{fig:clip-alignment-fpr}
    \end{minipage}
\end{figure}

\begin{table}[h]
    \centering
    \resizebox{0.45\textwidth}{!}{
        \begin{tabular}{cccc|cc}
            \toprule
            \multirow{2}{*}{\textbf{Model}} & \multicolumn{3}{c|}{\textbf{Model Details}} & \multicolumn{2}{c}{\textbf{Training Details}} \\ \cmidrule(lr){2-6} 
            & \textbf{\begin{tabular}[c]{@{}c@{}}Generated \\ Image Size\end{tabular}} & \textbf{Architecture} & \textbf{\begin{tabular}[c]{@{}c@{}}\# of \\ Parameters\end{tabular}} & \textbf{\begin{tabular}[c]{@{}c@{}}Batch\\ Size\end{tabular}} & \textbf{\begin{tabular}[c]{@{}c@{}}Time to\\ Train\end{tabular}} \\ \midrule
            SD1.5~\cite{sd15} & 512px & UNET + 1TE & 860M & 512 & 10 days \\ \midrule
            SD2.1~\cite{sd21} & 768px & UNET + 1TE & 860M & 256 & 1.2 days \\
            SDXL~\cite{sdxl} & 1024px & UNET + 2TE & 2.6B & 128 & 5.2 days \\
            FLUX~\cite{flux} & 1024px & DiT + 2TE & 12B & 64 & 9.3 days \\ \bottomrule
        \end{tabular}
    }
    \vspace{0.4cm}
    \caption{Model and training details. ``TE'' stands for ``text encoder''. We use a lr $=$1e-4 for all models and a linear warmup (except FLUX). All models are trained for 5000 steps.}
    \label{tab:diffusion-model-details}
    \vspace{-0.15in}
\end{table}

\subsection{Details on Identifying Successful Mislabeling using CLIP Similarity}
\label{app:msr-threshold-ablation}
In \S\ref{subsec:eval-metrics},  one of the three criteria for a successfully mislabeled image is whether the caption generated on the adversarial image more closely aligns with the target image than the reference image, i.e., 
\begin{eqnarray}
  \Delta=\text{CLIP(caption, TargetImg) - CLIP(caption, ReferenceImg)}>0 \nonumber
\end{eqnarray}
Here the choice of $0$ as the difference threshold is driven by the following empirical observations.

\para{Very few adversarial images have $\Delta \approx 0$.}  For an adversarial image, 
$\Delta= 0$ means the image captures a roughly equal mix of both reference and target image content. In this case, the mislabel attack is already partially successful. Across all of our experiments, we observe very few ($\approx 1$\%)  adversarial images with $\Delta \in [-5,5]$.

\para{Most adversarial images are precisely mislabeled.}
For a benign image, its generated caption is just the caption of the reference image, and we observe $\Delta\approx -27$ across all the benign images considered in our experiments.  Similarly, a perfect attack means $\Delta\approx 27$. In our experiments, 90\% of adversarial images have $\Delta>15$ and ~75\% have $\Delta>20$. This shows that a very high percentage of our mislabeled images are very precisely mislabeled to their target concept. 

Based on these two observations, we see that setting the difference threshold to 0 (i.e., $\Delta>0$) is sufficient,   although a stronger, more precise attacker could set a higher difference threshold to achieve more precise mislabeling.  On the other hand, given the distinct difference between benign images ($\Delta\approx -27$) and adversarial images (97\% of images with $\Delta>5$, 90\% with $\Delta>15$), varying the difference threshold between 0 and 5 will not cause any visible difference in TPR/FPR.  Finally, we also manually inspected all the image/caption pairs whose $\Delta$ is in (5,10), and confirmed that these images are all successfully mislabeled, without any false positives.    Together, these results support the use of $\Delta>0$ as a key criterion for successfully mislabeled images.

\begin{figure}[ht]
    \centering
        \centering\includegraphics[width=0.9\columnwidth]{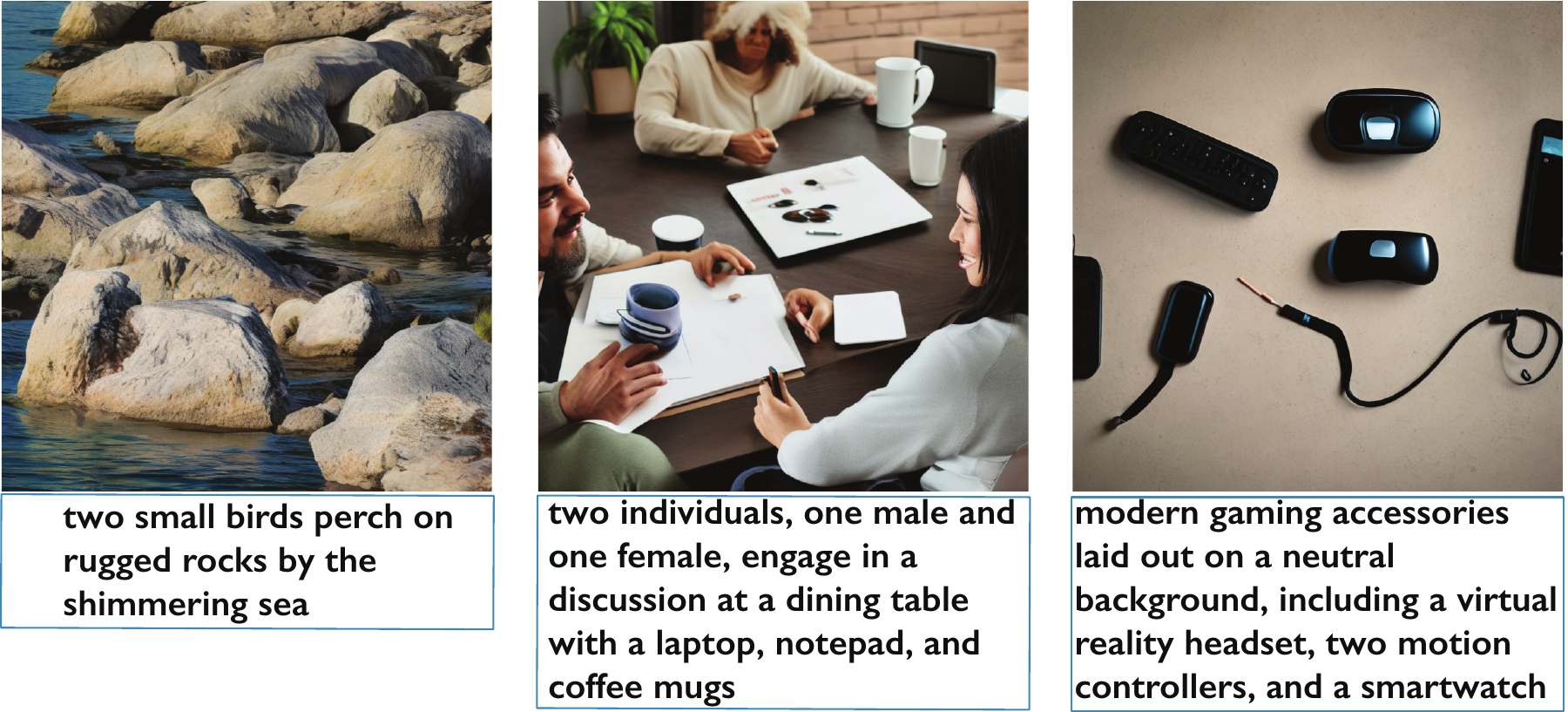}
        \caption{Images generated by SD2.1 fine-tuned on benign images whose captions were obtained after DiffPure.}
        \label{fig:diffpure-images}
        \vspace{-0.1in}
\end{figure}

\subsection{Image Generation after Countermeasures}
\label{app:image-gen-countermeasures}

We fine-tuned an SD2.1 model on 12,500 benign images with their CogVLM captions after applying DiffPure to the entire dataset. In Figure \ref{fig:diffpure-images}, we can see the generation quality with respect to the prompts is rather low. In the first image, with the prompt ``two small birds perch on rugged rocks by the shimmering sea,'' there are no birds. In the second image, with the prompt ``two individuals, one male and one female, engage in a discussion at a dining table with a laptop, notepad, and coffee mugs,'' the objects in the image are mixed, and there is an extra person. In the third image, with the prompt ``modern gaming accessories laid out on a neutral background, including a virtual reality headset, two motion controllers, and a smartwatch,'' there is a general lack of detail, making it difficult to understand what each object is supposed to be.

\subsection{CLIP Models for Black-Box Optimization}
\label{app:clip-models}
We select all four OpenAI CLIP variants (ViT-B-32, ViT-B-16, ViT-L-14, ViT-L-14-336) due to their popularity, as well as the current top four performing CLIP models (ViT-H-14-378-quickgelu, EVA02-E-14-plus, ViT-SO400M-14-SigLIP-384, ViT-bigG-14-CLIPA-336) with non-overlapping training datasets.